\documentclass[twocolumn, twocolappendix]{aastex631}

\usepackage{verbatim}
\usepackage[utf8]{inputenc}
\usepackage{cellspace}
\usepackage{amssymb}
\usepackage{subfigure}
\usepackage{amsmath}

\newcommand{\Xio}{\ensuremath{\xi_{\mathrm{ion,0}}}}
\newcommand{\Ha}{\ensuremath{\mathrm{H}\alpha}}
\newcommand{\Hb}{\ensuremath{\mathrm{H}\beta}}
\renewcommand{\textbf}[1]{#1}

\newcommand{\logxio}{\ensuremath{\log(\xi_{\rm ion, 0})}}

\begin{document}
\title{REBELS-IFU: Spatially Resolved Ionizing Photon Production Efficiencies of 12 Bright Galaxies in the Epoch of Reionization}

\author[0000-0002-5235-7971]{Lena Komarova}\email{olena.komarova@uv.es}
\affiliation{Departament d'Astronomia i Astrof\`isica, Universitat de Val\`encia, C. Dr. Moliner 50, E-46100 Burjassot, Val\`encia,  Spain}

\author[0000-0001-7768-5309]{Mauro Stefanon}
\affiliation{Departament d'Astronomia i Astrof\`isica, Universitat de Val\`encia, C. Dr. Moliner 50, E-46100 Burjassot, Val\`encia,  Spain}
\affiliation{Unidad Asociada CSIC "Grupo de Astrof\'isica Extragal\'actica y Cosmolog\'ia" (Instituto de F\'isica de Cantabria - Universitat de Val\`encia)}

\author[0000-0003-3316-3044]{Andrés Laza Ramos }
\affiliation{Departament d'Astronomia i Astrof\`isica, Universitat de Val\`encia, C. Dr. Moliner 50, E-46100 Burjassot, Val\`encia,  Spain}

\author[0000-0002-4205-9567]{Hiddo S.B. Algera}
\affiliation{Institute of Astronomy and Astrophysics, Academia Sinica, 11F of Astronomy-Mathematics Building, No.1, Sec. 4, Roosevelt Rd, Taipei 106319, Taiwan, R.O.C.}

\author[0000-0002-6290-3198]{Manuel Aravena}
\affiliation{Instituto de Estudios Astrof\'{\i}cos, Facultad de Ingenier\'{\i}a y Ciencias, Universidad Diego Portales, Av. Ej\'ercito 441, Santiago, Chile}
\affiliation{Millenium Nucleus for Galaxies (MINGAL)}

\author[0000-0002-4989-2471]{Rychard Bouwens}
\affiliation{Leiden Observatory, Leiden University, P.O. Box 9513, 2300 RA Leiden, The Netherlands}

\author[0000-0003-3917-1678]{Rebecca Bowler}
\affiliation{Jodrell Bank Centre for Astrophysics, Department of Physics and Astronomy, School of Natural Sciences, The University of Manchester, Manchester, M13 9PL, UK}

\author[0000-0001-9759-4797]{Elisabete da Cunha}
\affiliation{International Centre for Radio Astronomy Research, University of Western Australia, 35 Stirling Hwy., Crawley, WA 6009, Australia}

\author[0000-0001-8460-1564]{Pratika Dayal}
\affiliation{Canadian Institute for Theoretical Astrophysics, 60 St George St, University of Toronto, Toronto, ON M5S 3H8, Canada}
\affiliation{David A. Dunlap Department of Astronomy and Astrophysics, University of Toronto, 50 St George St, Toronto ON M5S 3H4, Canada}
\affiliation{Department of Physics, 60 St George St, University of Toronto, Toronto, ON M5S 3H8, Canada}

\author[0000-0002-9400-7312]{Andrea Ferrara}
\affiliation{Scuola Normale Superiore, Piazza dei Cavalieri 7, 56126, Pisa, Italy}

\author[0009-0008-3946-0502]{Rebecca Fisher}
\affiliation{Jodrell Bank Centre for Astrophysics, Department of Physics and Astronomy, School of Natural Sciences, The University of Manchester, Manchester, M13 9PL, UK}

\author[0000-0003-2804-0648]{Themiya Nanayakkara}
\affiliation{Centre for Astrophysics and Supercomputing, Swinburne University of Technology, P.O. Box 218, Hawthorn, 3122, VIC, Australia}

\author[0009-0009-2671-4160]{Lucie E. Rowland}
\affiliation{Leiden Observatory, Leiden University, P.O. Box 9513, 2300 RA Leiden, The Netherlands}

\author[0000-0001-9746-0924]{Sander Schouws}
\affiliation{Departament d'Astronomia i Astrof\`isica, Universitat de Val\`encia, C. Dr. Moliner 50, E-46100 Burjassot, Val\`encia,  Spain}
\affiliation{Leiden Observatory, Leiden University, P.O. Box 9513, 2300 RA Leiden, The Netherlands}

\author[0000-0001-8034-7802]{Renske Smit}
\affiliation{Astrophysics Research Institute, Liverpool John Moores University, 146 Brownlow Hill, Liverpool L3 5RF, United Kingdom}

\author[0000-0002-2906-2200]{Laura Sommovigo}
\affiliation{Center for Computational Astrophysics, Flatiron Institute, 162 5th Avenue, New York, NY 10010, USA}

\author[0000-0001-6106-5172]{Daniel P. Stark}
\affiliation{Department of Astronomy, University of California, Berkeley, Berkeley, CA 94720, USA}

\author[0000-0001-5434-5942]{Paul van der Werf}
\affiliation{Leiden Observatory, Leiden University, P.O. Box 9513, 2300 RA Leiden, The Netherlands}

\begin{abstract}
Measuring the ionizing photon production efficiency \Xio\, -- the ratio of ionizing photon output rate $Q_{\rm H^0}$ to UV continuum luminosity $L_{\rm UV}$ -- in galaxies at $z > 6$ is crucial for constraining their contribution to cosmic reionization. We present integrated and spatially resolved measurements of \Xio\ for 12 exceptionally bright ($M_\mathrm{UV} \sim -22$~mag) star-forming galaxies at $z \sim 7$ from the REBELS survey. These measurements are based on JWST NIRSpec/IFU PRISM spectroscopy, probing the rest-frame UV and optical regime. Notably, in 8 of the 12 galaxies, the spectral coverage includes H$\alpha$, enabling self-consistent dust attenuation estimates in both the ionized gas and stellar continuum via the Balmer decrement and rest-UV slope, respectively. We find global \logxio\ values ranging from $25.19\pm0.11$ to $25.61\pm0.11$, with a weighted mean of $25.44\pm0.15$, consistent with the canonical value of $\sim25.3$.
Using a sample of 25 star-forming clumps within these galaxies, we explore local variations in LyC production efficiency, finding a broader range, from $24.52\pm0.21$ to $26.18\pm0.61$. We identify strong correlations between \Xio\ and specific star formation rate, star formation surface density, H$\beta$ equivalent width, and stellar mass. Clumps with the highest \Xio\ exhibit $\rm EW_0(H\beta)$ $\geq150$~\AA, consistent with young stellar ages. From previous Ly$\alpha$ measurements in three galaxies, we estimate a typical Ly$\alpha$ escape fraction of $f_{\rm esc, Ly\alpha} \sim 2\%$, suggesting similar or lower escape fractions for LyC photons. Combining this with our \Ha\ measurements, we infer ionized bubble sizes $\sim 1$~pMpc, aligned with expectations from Ly$\alpha$-detected systems and reionization models.

\end{abstract}


\section{Introduction}
Characterizing the ionizing radiation of the earliest galaxies is key to our understanding of the Universe's history, as well as galaxy formation and evolution. It is of particular importance in the study of the epoch of reionization (EoR, $z \geq 6$), the last major phase transition of the Universe from a neutral to an ionized state \citep[e.g.,][]{Dayal2018}. The primary agents of reionization are thought to be star-forming galaxies \citep[e.g.,][]{Robertson2013, Robertson2015, Robertson2022rev, Bouwens2015, Finkelstein2019, Trebitsch2021, Naidu2020, Matthee2022, Matthee2023, Mascia2023, Yeh2023, Atek2024, Simmonds2024a, Dayal2025}, with active galactic nuclei (AGN) likely playing a minor role \citep{Duncan2015, Madau2015, Cristiani2016, Matsuoka2018, Grazian2024}. However, the relative contributions of numerous, low-mass galaxies versus rare, high-mass galaxies to reionization remain unclear \citep{Finkelstein2019, Naidu2020, Robertson2022}. In turn, the nature of the ionizing sources influences the timing and topology of reionization, where numerous low-mass galaxies may begin a gradual reionization earlier, while rare massive galaxies would instead drive a later, more rapid reionization via larger bubbles \citep[e.g.,][]{Lu2024, Hutter2021}. Constraining the ionizing properties of high-redshift galaxies across the mass spectrum is thus necessary to advance our understanding of reionization.

One of the key parameters in the physical models of reionization is the ionizing photon production efficiency, \Xio, defined as
\begin{equation}
    \Xio\ = \frac{Q_{H^0}}{L_{UV}},
\label{eqn:Xiion}
\end{equation}
where $Q_{\rm H^0}$ is the rate of hydrogen-ionizing photon ($\lambda < 912$~\AA) production, and $L_{UV}$ is the rest-frame UV continuum luminosity density. The subscript ``0'' in \Xio\ denotes the assumption of zero ionizing-radiation escape, i.e. a Lyman continuum escape fraction of $f_{\rm esc, LyC} = 0$. In reionization models, \Xio\ determines the emission rate of ionizing photons into the IGM, $\dot{n}_{\rm ion}$, from a galaxy population making up a UV luminosity density $\rho_{\rm UV}$ and with the LyC escape fraction $f_{\rm esc, LyC}$ as: $\dot{n}_{\rm ion} = \xi_{\rm ion}\times\rho_{\rm UV}\times f_{\rm esc, LyC}$. This is the key parameterization of the LyC emissivity of galaxies, which is not directly detectable in the EoR due to the opaque, neutral IGM \citep[e.g.,][]{Madau1995, Fan2001, McGreer2015, Inoue2014}. Thus, \Xio\ relates the observable, non-ionizing UV continuum of galaxies to their intrinsic ionizing continuum luminosity. Physically, \Xio\ traces the shape of the UV spectrum of star-forming galaxies below and above the Lyman limit, and represents the number of ionizing photons produced in a given star-forming system, with the assumption of no LyC escape. It is therefore a fundamental parameter in quantifying the ionizing photon budget available for reionization.

\begin{deluxetable*}{lcccc}
\tabletypesize{\small}
\tablecaption{REBELS‑IFU (JWST/NIRSpec) Galaxies}
\label{table:samp_props}
\tablehead{
\colhead{ID} & \colhead{$z$ \tablenotemark{\rm \scriptsize a}
} & \colhead{$M_{\rm UV}$ \tablenotemark{\rm \scriptsize b}
} & \colhead{$\beta$ \tablenotemark{\rm \scriptsize c}
} & \colhead{$\rm SFR_{\rm H\beta}~(\rm M_{\odot}~yr^{-1})$ \tablenotemark{\rm \scriptsize d}
}}
\startdata
REBELS-05 & $6.496$ & $-21.49 \pm 0.07$ & $-1.42 \pm 0.06$ & $126 \pm 54$ \\
REBELS-08 & $6.749$ & $-21.88 \pm 0.03$ & $-1.92 \pm 0.05$ & $133 \pm 79$ \\
REBELS-12 & $7.349$ & $-22.39 \pm 0.03$ & $-1.67 \pm 0.03$ & $101 \pm 124$ \\
REBELS-14 & $7.084$ & $-22.30 \pm 0.04$ & $-1.74 \pm 0.03$ & $159 \pm 42$ \\
REBELS-15 & $6.880$ & $-22.40 \pm 0.03$ & $-2.01 \pm 0.03$ & $207 \pm 74$ \\
REBELS-18 & $7.675$ & $-22.11 \pm 0.02$ & $-1.56 \pm 0.03$ & $76 \pm 22$ \\
REBELS-25 & $7.306$ & $-21.46 \pm 0.05$ & $-1.61 \pm 0.09$ & $73 \pm 25$ \\
REBELS-29 & $6.685$ & $-22.00 \pm 0.04$ & $-1.89 \pm 0.05$ & $64 \pm 28$ \\
REBELS-32 & $6.729$ & $-21.16 \pm 0.08$ & $-1.34 \pm 0.07$ & $103 \pm 39$ \\
REBELS-34 & $6.633$ & $-22.25 \pm 0.02$ & $-2.23 \pm 0.03$ & $63 \pm 61$ \\
REBELS-38 & $6.577$ & $-21.99 \pm 0.05$ & $-1.63 \pm 0.06$ & $182 \pm 115$ \\
REBELS-39 & $6.847$ & $-22.39 \pm 0.04$ & $-2.07 \pm 0.04$ & $156 \pm 63$ \\
\enddata
\tablenotetext{\rm a}{Spectroscopic redshift from the ALMA [C~II] detection \citep{Bouwens2022}.}
\tablenotetext{\rm b}{Rest-UV absolute magnitude, measured from NIRSpec IFU spectra with a top hat filter in the range $1450-1750$~\AA~by \cite{Fisher2025}.}
\tablenotetext{\rm c}{Rest-UV continuum slope, fitted in the $1268-2580~\text{\AA}$ range of the NIRSpec IFU spectra by \cite{Fisher2025}. }
\tablenotetext{\rm d}{Star formation rate, derived from the de-reddened \Hb\ luminosity, using the \cite{Kennicutt1998} conversion $\rm SFR_{\rm H\beta} = 5.5\times10^{-42}\times L_{H\beta}~(erg~s^{-1}) \times 2.86$, and assuming a \cite{Kroupa2001} IMF. The \Hb\ flux is adopted from \cite{Rowland2025}, and the dust correction is applied as detailed in Section~\ref{sec:analysis}.}
\end{deluxetable*}

The canonical value of \Xio\ adopted in reionization models is $\log(\Xio\ /(\rm Hz~erg^{-1})) = 25.2-25.3$ \citep[e.g,][]{Robertson2013, Wilkins2016}. This value is based on standard stellar population synthesis models assuming continuous star formation over $\sim 100$~Myr, near-solar metallicity, a Salpeter or Chabrier IMF ($0.1-100 \rm~M_{\odot}$), and no escape of ionizing photons. However, the value of \Xio\ is predicted to strongly depend on the stellar metallicity, initial mass function, and the star formation history, as well as stellar binary effects \citep[e.g.,][]{Zackrisson2013, Eldridge2017, Stanway2016, Stanway2018, Mauerhofer2025}. 

\textbf{Photometric and spectroscopic studies probing a large range of redshifts indicate an evolution of \Xio\,, where at increasingly earlier cosmic times, star-forming galaxies produce LyC photons more efficiently. The constraints on the redshift evolution of \Xio\ are drawn from heterogeneous galaxy samples, ranging from UV-complete selections to more targeted populations, such as extreme emission line galaxies (EELGs) and Ly$\alpha$ emitters. Despite these differences, studies generally suggest an increase of \Xio\ with increasing redshift, as briefly summarized below. Among broadly selected galaxy populations, UV-complete samples at $z \sim 2$ yield $\log(\Xio\ /(\rm Hz~erg^{-1})) \sim 25.5$ \citep{Emami2020}. At $z = 4-5$, values of $25.5-25.8$ are reported in sub-$L_*$ populations \citep{Bouwens2016, Lam2019}. Intriguingly, in Lyman Break galaxies (LBGs) at $z = 8$, $\log(\Xio\ /(\rm Hz~erg^{-1}))$ values appear to reach $25.8-26.0$ \citep{DeBarros2019, Stefanon2022}, nearly an order of magnitude higher than canonically assumed. At the highest currently probed redshifts ($z \sim 10-12$), a handful of galaxies show \Xio = $25.3-25.7$, possibly indicating a flattening of \Xio$(z)$ \citep{Hsiao2024, Calabro2024, AlvarezMarquez2025}. Studies aimed specifically at investigating the redshift evolution of \Xio\ indeed find it to increase with redshift, e.g. from $z = 4-9$ \citep{Llerena2024_Xi, Simmonds2024}. As for the more narrowly selected galaxy samples, low-redshift, $z \sim 0.3$, EoR galaxy analogs with direct LyC observations show \Xio\ consistent with the canonically assumed values \citep{Schaerer2016}. But by $z \sim 3.5$, strong emission-line galaxies are reported to have $\log(\Xio\ /(\rm Hz~erg^{-1}))$ as high as 25.8 \citep{Nakajima2016, Onodera2020}. In a large sample of $z = 4-9$ EELGs, the majority show \Xio\ values higher than canonical \citep{Llerena2024CEERS}. Similarly, Ly$\alpha$ emitters at $z = 4.0 - 13.4$ appear to fall at \Xio $ > 25.5$ \citep{Heintz2025}, with some extreme cases extending to 26.3 \citep{Maseda2020}. However, it is worth noting that e.g., \cite{Castellano2023} do not find \Xio\ to change significantly with redshift in $z = 2-5$ massive galaxies. Similarly, samples of EELGs at $1.3 < z < 4.0$ show moderate median values of $25.2$ \citep{Tang2019, Jaiswar2024}}


The value and redshift evolution of \Xio\,, in turn, has important implications for reionization models \citep{Stark2025}. In particular, the values of \Xio\ revealed by JWST observations may suggest an overabundance of ionizing photons, which in turn would lead to reionization ending earlier than expected from cosmic microwave background and Ly$\alpha$ forest constraints \citep{Munoz2024}. Although, this problem may be resolved if high \Xio\ values are accompanied by low LyC escape fractions \citep{Matthee2023, Atek2024, Papovich2025}. Another possible solution is that a substantial population of faint, weak-line galaxies uncovered by JWST at $z > 7$ have declining star formation histories, and this post-burst phase leads to a broadening of the \Xio\ distribution to lower values \citep{Endsley2024}. \Xio\ is thus a key parameter to constrain in EoR galaxies. 

In this paper, we present spatially resolved measurements of \Xio\ in reionization-era massive galaxies from the REBELS program. With JWST/NIRSpec-IFU rest-UV to optical spectra of twelve UV-bright galaxies, we resolve the ionizing photon production efficiency during the EoR on $2-3$~kpc scales, and relate it to global and local physical properties and conditions. 

We describe our sample and observations in Section~\ref{sec:sample_data}, and our analysis in Section~\ref{sec:analysis}. We present our results in Section~\ref{sec:results}, discuss them in Section~\ref{sec:discussion}, and summarize our conclusions in Section~\ref{sec:conclusions}. 
Throughout this work, we adopt magnitudes in the AB system \citep{Oke1983}, a \cite{Kroupa2001} IMF, and a $\Lambda$CDM cosmology with $\Omega_\Lambda = 0.7$, $\Omega_M = 0.3$, and $H_0 = 70~\mathrm{km\,s^{-1}\,Mpc^{-1}}$, consistent with \cite{Planck2020}. All \Xio\ measurements are expressed in logarithmic units of $\rm Hz~erg^{-1}$, unless otherwise specified.

\section{Sample and Observations}
\label{sec:sample_data}
\subsection{Galaxy Sample}

We investigate a sub-sample of 12 galaxies from the Cycle 7 ALMA Large Program REBELS -- Reionization Era Bright Emission Line Survey 
\citep{Bouwens2022} and its pilot programs \citep{Smit2018, Inami2022, Schouws2022, Schouws2023}. The galaxies in the REBELS program were initially photometrically selected using standard Lyman-break criteria in deep ground-based, wide-field imaging \citep{Bouwens2022}. \textbf{They currently constitute the largest sample of bright UV-selected galaxies at $z\sim7$. Several additional bright candidates have been identified in JWST fields, largely without ALMA follow-up \citep[e.g.,][]{Roberts-Borsani2025, Jung2022, Finkelstein2022, Tang2023}. Because REBELS relies on conventional LBG selection, the sample is likely representative of the bright, star-forming, moderately dusty, LBG population. However, quantifying how well it represents the broader UV-luminous population at $z\geq6.5$ requires a mass-complete sample with ALMA coverage in the dust continuum at these redshifts and luminosities, which is not yet available.} 

Employing a spectral scan technique targeting [C~II]$158~\rm \mu m$ and dust continuum emission, the REBELS survey is designed to characterize the star formation, dust, and ISM conditions in the most UV-luminous star-forming galaxies at $z_{\rm phot} > 6.5$. \textbf{ALMA measurements indicate that REBELS galaxies are substantially dust-enriched, showing dust-to-gas ratios similar to those of local metal-rich systems, as well as obscured star-formation fractions of $50-80\%$ \citep{Inami2022, Ferrara2022, Sommovigo2022, Algera2025, Fisher2025b}. However, their massive dust reservoirs may not be unique in the EoR, as the dust-to-gas and dust-to-stellar mass ratios in REBELS are broadly consistent with galaxies of similar redshift and metallicity \citep{Algera2025}. }

Twelve of the most luminous [C~II]$158~\rm \mu m$ emitters in the REBELS program were selected for follow-up spatially resolved spectroscopy with the JWST NIRSpec Integral Field Unit (IFU), covering rest-UV to optical wavelengths. The galaxies in this REBELS-IFU sub-sample ($6.5 \leq z \leq 7.7$) were chosen to evenly sample the parameter space of stellar mass, rest-frame UV continuum slope $\beta$, and [O~III]+\Hb\ equivalent width, and are therefore representative of the larger REBELS sample (Stefanon et al. in prep), though with a bias towards [C~II]- and dust-luminous galaxies. \textbf{REBELS-IFU galaxies are exceptionally bright, with rest-frame UV magnitudes $M_{\rm UV} = -22.7 \rm~to -21.6 $ and high stellar masses $\log(M_{*,\rm tot}/\rm M_{\odot}) = 9.2-9.7$.} As shown in Table~\ref{table:samp_props}, these galaxies span a range of observed physical properties, i.e. rest-frame UV continuum slopes $\beta = -2.05~\rm to -1.2$, metallicities $\log(\rm O/H) + 12 = 7.8-8.7$, star formation rates $\rm SFR_{H\beta} = 64-204 \rm~M_{\odot}~yr^{-1}$, and $\rm O_{32} = 1.9-9.4$ \citep[][Stefanon et al. in prep]{Bouwens2022, Rowland2025, Fisher2025}. 

\subsection{Observations}
\label{sec:obs}
The JWST NIRSpec IFU prism observations for all targets except REBELS-18 were obtained through the Cycle 1 program GO 1626 (PI: Stefanon). For REBELS-18, the IFU prism coverage included in GO 1626 overlapped with that of program GO 2659 (PI: Weaver), and the observations were carried out on a shared basis. The NIRSpec IFU covers a $3\arcsec \times 3\arcsec$ field of view, with our adopted spatial sampling of $0.08\arcsec$/pixel. The prism configuration of the NIRSpec IFU provides a nominal resolving power of $R \sim 100$ in the wavelength range $0.6-5.3~\rm \mu m$. In our sample, it therefore covers rest-UV to optical wavelengths, including the [O~II]$\lambda\lambda$3727,3729, \Hb\, and [O~III]$\lambda\lambda$4959,5007 emission lines. Moreover, \Ha\ is covered for the eight galaxies at $z \leq 7$, while for the remaining four, it is redshifted out of NIRSpec coverage. All sources except REBELS-18 were observed with an exposure time of $\sim 30$~minutes, sufficient for robust (SNR $\geq10$) continuum and line detections. In contrast, REBELS-18 was observed for $\sim 95$~minutes as part of GO 2659. Data reduction was performed with the JWST Science Calibration Pipeline. For galaxy-integrated measurements, a combined spectrum was extracted for each target from an aperture defined by merging $7\sigma$ isophotes in multiple key bands, covering UV and optical continuum, as well as emission lines. A complete description of the observational setup, data reduction, and spectral extraction procedure is detailed by Stefanon et al. in prep. 

\section{Analysis}
\label{sec:analysis}
Following the procedure by \cite{Bouwens2016}, we compute the ionizing photon production efficiency \Xio\ (Eq. \ref{eqn:Xiion}) from the ratio of the intrinsic \Ha\ and rest-frame $1500$~\AA~continuum luminosities $L_{\rm H\alpha}$ and $L_{1500}$, respectively. The de-reddened \Ha\ luminosity traces the rate of ionizing photon production $Q_{\rm H^0}$, assuming Case~B recombination in ionization‑bounded H~II regions with no LyC escape: $L_{\rm H\alpha}~(\rm erg~s^{-1}) = 1.36 \times 10^{-12}~\textit{Q}_{H^0}~(\rm s^{-1})$ \citep{Leitherer1995, Kennicutt1998}. However, we caution that shocks, outflows, or density‑bounded leakage that break these assumptions could bias \Xio\ measurements. For the four objects at $z \geq 7$ REBELS-12, 14, 18, and 25, \Ha\ is redshifted out of NIRSpec coverage, and we thus instead use the dust-corrected \Hb\ luminosity, using an intrinsic ratio of \Ha/\Hb\ $= 2.86$, assuming Case B recombination. 
$Q_{\rm H^0}$ is then normalized by the attenuation-corrected UV luminosity to obtain \Xio.  Each measurement of \Xio\ thus requires a measurement of $L_{\rm H\alpha}$, $L_{\rm H\beta}$, $L_{1500}$, and UV continuum slope $\beta$ for dust correction. We perform these measurements on two separate spatial scales: on the global, galaxy-integrated scale, and on the scale of individual star-forming clumps (Section~\ref{sec:spatialbinning}) within each galaxy. To compute \Xio\ in clumps, we use the measurements of UV continuum luminosity and slope, as well as line luminosities, within each clump. Below, we detail the procedure used to derive these parameters for each galaxy.

Emission line measurements of the integrated NIRSpec IFU spectra have been presented by \cite{Rowland2025}. In this work, we also aim to perform a spatially-resolved analysis, measuring the line intensities in sub-galactic structures. We therefore use the galaxy-integrated line measurements in \cite{Rowland2025} and adopt their procedure for spatially resolved measurements. To obtain the line luminosities, we model the optical continuum as a third-order polynomial, and fit the following emission lines with Gaussian profiles: \Ha\, \Hb, $[\rm O\textsc{iii}$]$\lambda\lambda$4959,5007, $[\rm O\textsc{ii}$]$\lambda\lambda$3727,3729, H$\epsilon$, H$\delta$, H$\gamma$+[O~III]$\lambda$4363, and $[\rm Ne\textsc{iii}$]$\lambda$3869. In galaxy-integrated and clump-integrated spectra (see below), we deblend \Ha\ from the [N~II]$\lambda\lambda$6548,6584 doublet by simultaneously fitting three Gaussians, with the amplitude ratio [N~II]$\lambda$6584/[N~II]$\lambda$6548 fixed to 3.049 \citep{Dojvcinovic2023}. On the other hand, in most individual spaxels, the signal-to-noise ratio is too low to deblend [N~II], and we thus  forgo this step when creating \Ha\ images for clump identification (Section~\ref{sec:spatialbinning}). This does not significantly affect the resulting image morphology, since the average [N~II]/(\Ha+[N~II]) $\sim 17\%$ in integrated spectra. All the considered emission lines are not spectrally resolved, and we fix their widths to the wavelength-dependent line spread function modeled for this dataset (Stefanon et al. in prep). We perform the line fitting using the least squares minimizer in the Python package {\tt lmfit}. The line fluxes are then computed from the best-fit line models, corrected for local continuum.

The galaxy-integrated rest-UV continuum luminosities $L_{1500}$ and slopes $\beta$ have been measured in the REBELS-IFU sample by \cite{Fisher2025}. We adopt their values for the global \Xio\ measurements and implement their procedure in the spatially resolved analysis. In particular, $L_{1500}$ is computed as the mean luminosity density in the range $\lambda_{\rm rest}  = 1450-1550$~\AA, and the slope $\beta$ is obtained from a power-law fit to the flux in the range $1268~\text{\AA} \leq \lambda_{\rm rest} \leq 2580~\text{\AA}$. In spatially resolved spectra, we bin this segment into 30 wavelength bins of width $45~\text{\AA}$ prior to fitting, to increase the signal-to-noise ratio (SNR).
We have tested this binning approach with the integrated spectra, and found that the slopes are recovered with $\leq 0.5\sigma$ accuracy.

\begin{deluxetable*}{lllll}
\tablecaption{{\tt BAGPIPES} SED model parameters and priors \label{tab:priors}}
\tablehead{
\colhead{Component} & \colhead{Parameter} & \colhead{Symbol / Unit} & \colhead{Range} & \colhead{Prior}
}
\startdata
Global & Redshift & $z$ & $z_{\rm spec}$ & -- \\
Dust  & Attenuation & $A_V$ & (0, 4) & Uniform \\
  & Deviation from \cite{Calzetti2000} & $\delta$ & F25 & Fixed \\
  & UV bump & B & F25 & Fixed \\
Nebular  & Ionization parameter & $\log(U)$ & (-3, -1) & Uniform \\
SFH  & Stellar mass formed & log($M^{\star}/M_{\odot}$) & (7, 12) & Uniform \\
 & Metallicity & $Z/Z_{\odot}$ & (0.01, 3) & Gaussian \\
 & Bins & -- & 4 & Student's t \\
\enddata
\textbf{Notes.} 
The custom dust attenuation curve parameters are adopted from \cite{Fisher2025} (F25).
\end{deluxetable*}

For the dust extinction correction, we adopt the custom attenuation laws derived by \cite{Fisher2025} for our galaxy sample,  based on SED fitting. These flexible attenuation curves turn out to often deviate from the \cite{Calzetti2000} law, and show significant UV bumps. We also adopt the global stellar continuum extinctions $\rm E(B-V)_{stellar}$ and intrinsic UV continuum slopes $\beta_0$ presented by \cite{Fisher2025}. The $\beta_0$ values they find are in the range $-2.05 \pm0.12$ to $-2.33\pm0.05$, consistent with the steepest slope we observe in our sample of star-forming clumps, $\beta_0 = -2.40 \pm0.10$. While the attenuation laws were derived for the galaxy-integrated spectra, we also adopt them, together with the $\beta_0$ values, for the spatially resolved measurements. However, to dust-correct the local UV continuum luminosities $L_{1500}$, we re-normalize the global attenuation curves using the spatially resolved, local UV continuum slopes $\beta$.

The emission line luminosities are corrected for dust extinction differently from the UV continuum. Nebular and stellar continuum dust attenuations are known to differ within a given galaxy, where the stellar optical depth is $\sim 0.4\times$ that of the gas, likely due to these components not being spatially coincident \citep[e.g.,][]{Calzetti1997, Kreckel2013, Reddy2015}. \cite{Fisher2025b} find an average $\rm E(B-V)_{\rm stellar}/E(B-V)_{\rm nebular} = 0.50$, with a scatter of 0.27, in our REBELS-IFU sample, consistent with previous measurements. We therefore adopt this ratio to estimate $E(B-V)_{\rm nebular}$ from $\beta$-based $E(B-V)_{\rm stellar}$ for the four galaxies at $z \geq 7.0$, for which the Balmer decrement cannot be measured. We then apply this nebular dust extinction correction to the \Hb\ luminosity, to obtain $Q_{\rm H^0}$. For the rest of the sample, we compute the nebular attenuation directly from the Balmer decrement, assuming Case~B recombination in a nebula of temperature $T = 10^4$~K and density $n_e = 100~\rm cm^{-3}$, corresponding to an intrinsic ratio \Ha/\Hb= 2.86 \citep{Osterbrock2006}. As a test of consistency, we have compared \Xio\ measurements from \Ha\ luminosity and the Balmer decrement for $z \leq 7.0$ galaxies, to those obtained from \Hb\,, $\beta$ and the 0.50 conversion factor above. The \Xio\ values are recovered within uncertainties.

To estimate the noise in the clump spectra, we compute the flux standard deviation in a running window of width $0.1~\rm \mu m$, after masking out emission lines. This results in $\sim 2\times$ higher uncertainties than those obtained from the error extension in the IFU data cube, assuming uncorrelated noise, similar to that found by \cite{Ubler2023}, \cite{Lamperti2024}, and Stefanon et al. in prep. For all of the above parameters, i.e., line fluxes, UV luminosity densities, and UV slopes $\beta$, we estimate the uncertainties via Monte Carlo sampling of the spectra $10^4$ times within these flux uncertainties. We propagate these uncertainties to the calculation of \Xio\ by re-sampling $10^4$ times our resulting distributions of \Ha\ and/or \Hb\ luminosities, UV luminosity density, and $\beta$.

We adopt the galaxy stellar masses presented by \cite{Fisher2025}, obtained with their flexible dust attenuation curves. We note that \cite{Fisher2025} report total formed stellar masses $M_{*,\rm tot}$, i.e. the integral of the star formation history, rather than the surviving stellar mass. We follow their approach, and all stellar masses quoted here refer to total formed mass. To estimate the spatially resolved stellar masses of clumps, we fit the clump NIRSpec IFU spectra with {\tt BAGPIPES} \citep{Carnall2018}. Nebular continuum and line emission parameters are generated using standard CLOUDY grids \citep{Ferland2017} in {\tt BAGPIPES}. We adopt a non-parametric star formation history (SFH), split into time bins of constant SFR. The non-parametric SFH model has been shown to likely mitigate the effects of overshining by young stellar populations \citep{Leja2019, Topping2022_rebels}. We use four time bins: two covering $0–3$~Myr and $3–7$~Myr before observation, and two spanning 10~Myr before observation to $z=30$, spaced uniformly in logarithmic time. We adopt the custom dust attenuation curves determined for our galaxies by \cite{Fisher2025}. Table~\ref{tab:priors} shows a summary of the parameter limits and the prior distributions used in our fitting.

\subsection{Clump Extraction}
\label{sec:spatialbinning}

\begin{figure*}
\includegraphics[width=\textwidth]{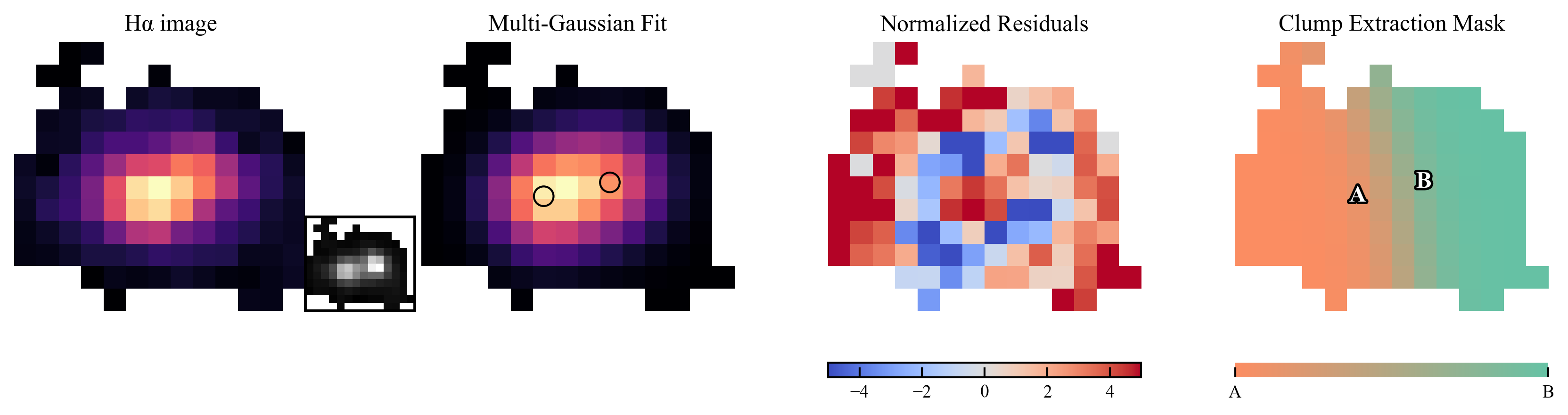} 
\caption{Illustration of the procedure used to determine the mask for extracting individual clumps, shown here for REBELS-15. From left to right, the panels show: continuum-subtracted \Ha\ flux, used for clump identification, within multi-wavelength $7\sigma$ isophotes (Section~\ref{sec:obs}; Stefanon et al. in prep), with an inset UV continuum image ($1250-2600$~\AA)}; 2D multi-Gaussian model fit to this image; residuals normalized by local noise level; the final mask used to extract the spectra of each clump from the IFU cube, with the color bar indicating the fractional flux contribution assigned to each region of the mask. The best-fit clump centroids are indicated with black circles in the second panel and labeled A and B in the last one. 
\label{fig:2dfits}
\end{figure*}

Spatially resolved measurements of \Xio\ cannot be performed on arbitrary spatial components, such as individual spaxels or SNR-based bins. This is because computing the rate of ionizing photon production $Q_{\rm H^0}$ from calibrations of recombination lines relies on the assumption of Case B recombination in an ionization-bounded region. We must therefore use spatial elements that can reasonably be treated as closed systems, where all locally produced ionizing photons are reprocessed. In our case, we spatially resolve \Xio\ on the scale of individual nebular clumps. 

To determine the locations and boundaries of the clumps, we fit the continuum-subtracted  \Ha\ (\Hb\ for $z \geq 7.0$) images with multiple 2D Gaussian components. We determine the number of clumps to fit in each galaxy based on a visual inspection of the nebular and UV continuum images. A preliminary cross-check with the new NIRCam imaging for a subset of our sample (PID 6480, PI: Schouws) confirms these identifications and, in some cases, reveals additional clumps. Setting the apparent emission peak locations as initial guesses, we perform a least-squares fit with the \texttt{scipy least\_squares} routine. Thus we obtain the best-fitting 2D emission models, from which we generate clump extraction masks that quantify, for each spaxel, the fractional contribution associated with each specific clump. We note that in this first-order analysis, we do not perform PSF matching for our clump extraction masks across wavelength. Finally, we use these masks to extract the spectrum of each clump from the IFU cubes. This process is illustrated in Figure~\ref{fig:2dfits} on one object, showing the emission-line image, the 2D model, and the extraction mask. The corresponding plots for all objects are presented in the appendix in Figure~\ref{fig:2dfits_app}. 

We note that Gaussian models are sufficient for our purposes of separating the galaxies into clumps, even though, as can be seen in the residual maps in Figures~\ref{fig:2dfits} and \ref{fig:2dfits_app}, they do not always account for the faint, diffuse emission at larger scales. This may be due to the presence of additional distinct emission components, or non-Gaussian nature of the compact clumps. The purpose of these models is only to disentangle the spectra of what can be considered individual nebular components, and we therefore simply assume that the residual diffuse emission in a given spaxel belongs to the Gaussian clump closest to it. Thus, for each galaxy, the sum of all of the clump-extracted spectra add up to the galaxy-integrated spectrum.

One caveat in this analysis is that we identify clumps in the nebular images without applying a dust correction to each spaxel. This may impact the resulting morphologies of the clumps, in particular in strongly obscured galaxies, such as REBELS-25 \citep{Rowland2024}. The spatially resolved dust morphologies of REBELS-IFU galaxies will be investigated in a future study. 

We find that the \Ha\ or \Hb\ emission in each REBELS-IFU galaxy can be decomposed into $1-3$ individual clumps, with a total of 25 clumps across 12 galaxies. We note that, because the \Hb\ image of REBELS-25 lacks the signal for 2D fitting, we manually select the clump in its northern edge, delineated by the global IFU extraction mask (Stefanon et al. in prep). Notably, this excludes the central, dust-obscured component of the galaxy \citep{Rowland2024} from the resolved analysis.

It is also important to note that REBELS galaxies have heterogeneous morphologies, and when we refer to the extracted 25 clumps, these include both star-forming regions within a given galaxy (e.g., REBELS-15, REBELS-18, REBELS-29), as well as clumps consisting of possibly independent, merging, lower-mass galaxies (e.g., REBELS-14, REBELS-39). These morphological differences can be seen in Figure~\ref{fig:2dfits_app} in Appendix~A.

\section{Results}
\label{sec:results}

\begin{figure*}
\includegraphics[width=0.9\textwidth]{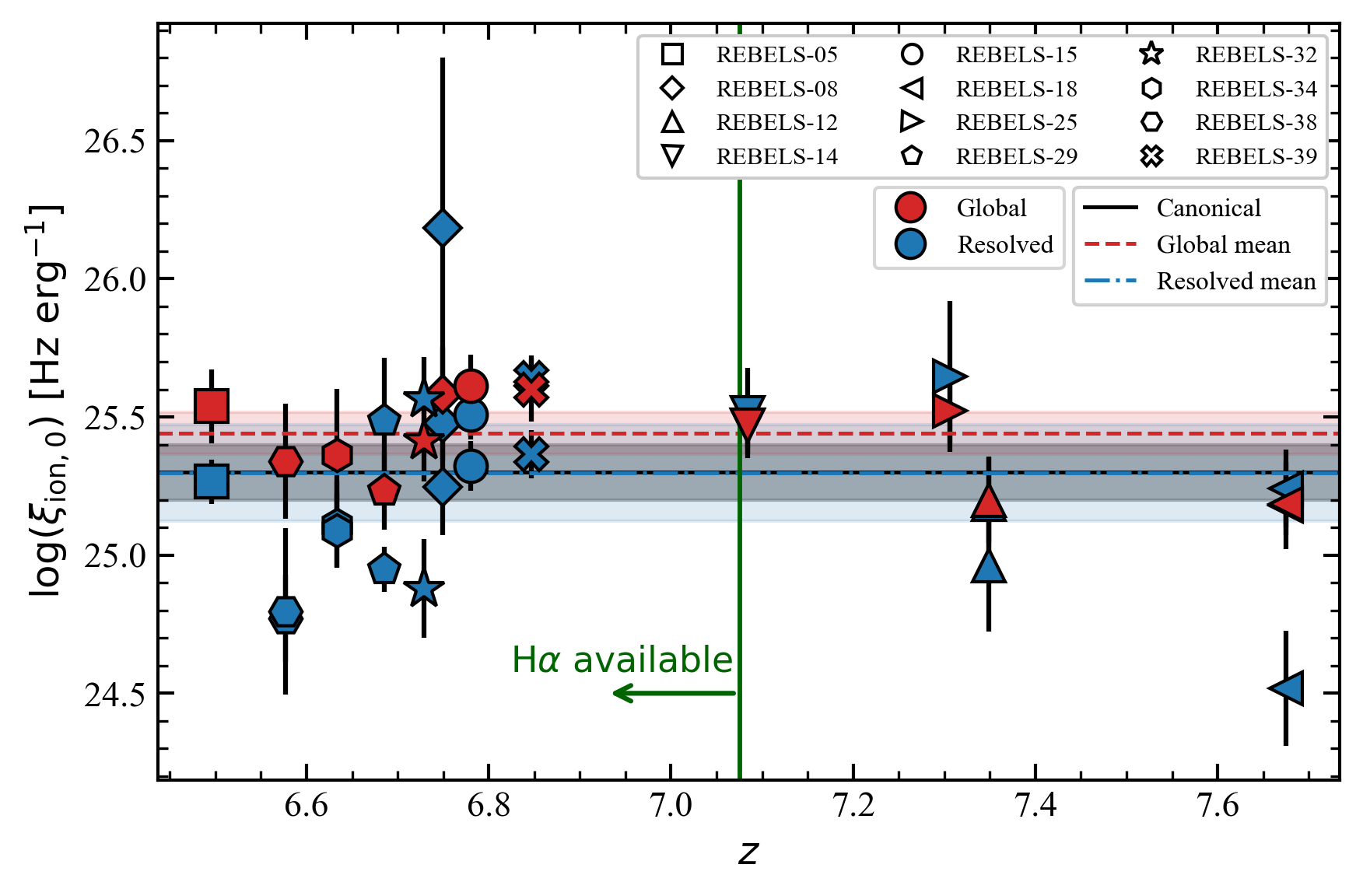} 
\caption{Ionizing photon production efficiencies \Xio\ in the REBELS-IFU sample vs. redshift. Each galaxy is plotted with a distinct marker symbol, as indicated in the legend. Global, galaxy-integrated values are shown in red, and the spatially resolved measurements in individual clumps are shown in blue. The average \Xio\ on each scale is shown as a horizontal band of the corresponding color, with its width representing the 1$\sigma$ dispersion of the measurements. The canonical value of \Xio\ \citep{Robertson2013, Wilkins2016} is marked as a black line. Measurements of \Xio\ present a large ($\geq 1$ dex) dispersion, but are on average consistent with the canonical value.}
\label{fig:global_resolved}
\end{figure*}

\subsection{Global and Resolved Measurements of \Xio}

We begin by presenting the global, galaxy-integrated values of \Xio\ for our sample, which provide a baseline measure of the ionizing photon production efficiency in REBELS galaxies. These measurements are presented in Table~\ref{table:globalXi} and shown in Figure~\ref{fig:global_resolved}.

We find that our \logxio\ values range from $25.19\pm0.11$ to $25.61\pm0.11$, with the sample average of $25.44\pm0.15$, where \Xio\ is in units of $\rm Hz~erg^{-1}$. All but two galaxies in our sample are within 2$\sigma$ of the canonical value of 25.3. 
The intriguing exceptions are REBELS-15 and REBELS-39, which exhibit \logxio\ values $\sim25.6 \pm 0.1$, or a factor of $\sim2$ higher than the canonical value of 25.3. We note that the dust correction changes the \logxio\ values by on average $\sim0.3$~dex and thus serves as a source of significant systematic uncertainty. 
In Section~\ref{subsec:dustcorr}, we discuss this in further detail.

Beyond the galaxy-integrated measurements, the JWST NIRSpec-IFU observations of our sample allow for an unprecedented view into individual clumps in EoR galaxies. The 25 clumps we extract from the \Ha\ or \Hb\ morphologies of our 12 galaxies have circularized radii $0.78-2.15$~kpc, computed as the radii of circles with the pixel area of each clump. The stellar masses of the clumps we obtain with SED fitting are in the range $\log(M_{*,\rm tot}/\rm M_{\odot}) = {7.70}-{9.71}$.

Our spatially resolved measurements of \Xio\ in these star-forming clumps are shown in Figure~\ref{fig:global_resolved} and listed in Table~\ref{table:resolvedXi}. The resolved analysis reveals a large range of \Xio\ across individual clumps, from $24.52\pm0.21$ to $26.18\pm0.61$. We find differences between global and resolved \Xio\ ranging from $-0.67$~dex to 0.60~dex, \textbf{i.e. resolved measurements can be either higher or lower than the global one.} \textbf{The reason for this is that \Xio\ is a property of individual stellar populations. When it is measured on global scales, the emission from all stellar populations is integrated and corrected using a single, galaxy-averaged dust attenuation. However, spatially resolved dust corrections applied to local luminosities more accurately trace the spectra of individual stellar populations. Together with the fact that \Ha\ and UV luminosities do not maintain a constant ratio across clumps, this leads to different \Xio\ values in global vs. resolved measurements.} On average, however, the resolved values are only 0.14~dex lower than the global ones. One interesting outlier is the clump REBELS-08-C, showing an extremely high \logxio = $26.19\pm0.61$. While consistent with the global \logxio\ of $25.58\pm0.17$ and the canonical value within $1\sigma$ and $1.5\sigma$, respectively, this result hints at the possible existence of star-forming regions in massive EoR galaxies that are unusually efficient producers of ionizing photons. 

\subsection{Caveat in $\xi_{\rm ion,0}$ Measurements: Dust Correction}
\label{subsec:dustcorr}
In estimating the ionizing photon production efficiency from the \Ha\ and UV continuum luminosities, the correction for dust extinction is paramount. For example, \cite{Schaerer2016} report that applying a dust correction lowers their estimates of \Xio\ by a factor of $2–6$. Most \Xio\ measurements in the literature assume uniform dust attenuation for the gas and stellar components, due to a lack of constraints on both. However, a multitude of studies suggests that in star-forming galaxies at $z \sim 0-3$, the stellar UV continuum extinction is $\sim 0.4\times$ the nebular gas extinction \citep{Calzetti1994, Calzetti2000, Mancini2011, Price2014, Reddy2015, Battisti2016, Qin2019, Shivaei2020}. This differential dust attenuation can be expected if the dust within the galaxies consists of a diffuse component, slightly attenuating all stellar emission, as well as dense clouds around young stars that additionally attenuate the nebular emission \citep[e.g.,][]{Calzetti1994, CharlotFall2000, Shivaei2020, Sommovigo2025}. In our measurements for objects without \Ha\ coverage, we have adopted the ratio $\rm E(B-V)_{\rm stellar}/E(B-V)_{\rm nebular} = 0.50$, with a scatter of 0.27, derived by \cite{Fisher2025b} for this sample. This conversion factor therefore introduces additional uncertainty in \Xio\ measurements for $z \geq 7.0$ objects. 

Overall, dust correction decreases our global \logxio\ values by up to $0.43$~dex, with a sample average of 0.18~dex.
As for the resolved measurements, the effect is even more significant, ranging from $0.00-0.89$~dex, on average 0.34~dex. In addition, we adopt the custom dust attenuation curves derived for each of our sources by \cite{Fisher2025}. Most of these curves are steeper than the \cite{Calzetti2000} law. If we instead adopted the \textbf{\cite{Calzetti2000} or the SMC extinction law, our global \logxio\ values would be on average 0.14~dex and 0.05~dex lower, respectively. }

\subsection{Relation of $\xi_{\rm ion,0}$\ to Physical Properties}

Significant work has been done on identifying the physical conditions and galaxy properties associated with a high ionizing photon production efficiency. It has been shown that more intense and recent star formation is the primary driver of high \Xio, as expected, since more OB star populations \textbf{increase the production of ionizing photons per UV luminosity}. Thus, correlations between \Xio\ and specific star formation rate, and stellar age tracers such as optical emission-line EWs, have been established \citep{Matthee2017,Chevallard2018,  Tang2019, Atek2022, Llerena2024_Xi, PrietoLyon2023, Castellano2023, Laseter2025}. These studies have also found \Xio\ to increase with lower stellar mass, lower metallicity, and fainter UV luminosity. In other words, low-mass, metal-poor, young line emitters are the most efficient producers of ionizing photons, while more massive, metal-rich galaxies are less efficient. Additionally, high \Xio\ is linked to higher $\rm O_{32}$ \citep{Llerena2024_Xi}, as expected, since harder radiation fields produce stronger ionization. It is also linked to higher star formation surface density \citep{Castellano2023}, where more dense star formation accompanies more efficient LyC production.

To investigate what drives the ionizing production efficiencies in our sample, we test correlations between \Xio\ and various parameters, both on the scale of galaxies and individual clumps when possible. In particular, for the integrated values, we investigate the relationship of \Xio\ to redshift; the observed rest-UV, \Ha\, and $\rm [C~II]_{158\mu m}$ luminosities; observed UV slope $\beta$; the rest-frame equivalent width of H$\beta$ $\rm EW_0(H\beta)$; specific star formation rate $\rm sSFR_{\rm H\beta}$, computed from de-reddened \Hb\ luminosities, using the \cite{Kennicutt1998} conversion factor; star formation rate surface density $\Sigma_{\mathrm{SFR, H\beta}}$; ionization parameter $\log(U)$ and oxygen abundance $12 + \log(\rm O/H)$, \textbf{derived from line ratios $O_{32}$ and $R_{23}$, respectively,} by \cite{Rowland2025} ; and stellar mass $M_{*,\rm tot}$. It is worth noting that some of the above quantities are not fully independent of each other. For instance, $\rm EW_0(H\beta)$, sSFR, and $\Sigma_{\mathrm{SFR, H\beta}}$ all depend on \Hb\ luminosity, and $\rm EW_0(H\beta)$ and sSFR are not independent because the optical continuum luminosity is proportional to the stellar mass. For the clump-scale correlations, we consider the same parameters, but excluding [C~II] luminosities due to the lack of resolved measurements. We also do not present clump-scale $\log(U)$ and $12 + \log(\rm O/H)$ measurements, \textbf{because, although all relevant emission lines are detected at sufficient SNR, the combined uncertainty from dust correction assumptions and line-ratio propagation leads to fractional uncertainties greater than unity.}.

\begin{figure*}
\includegraphics[width=\textwidth]{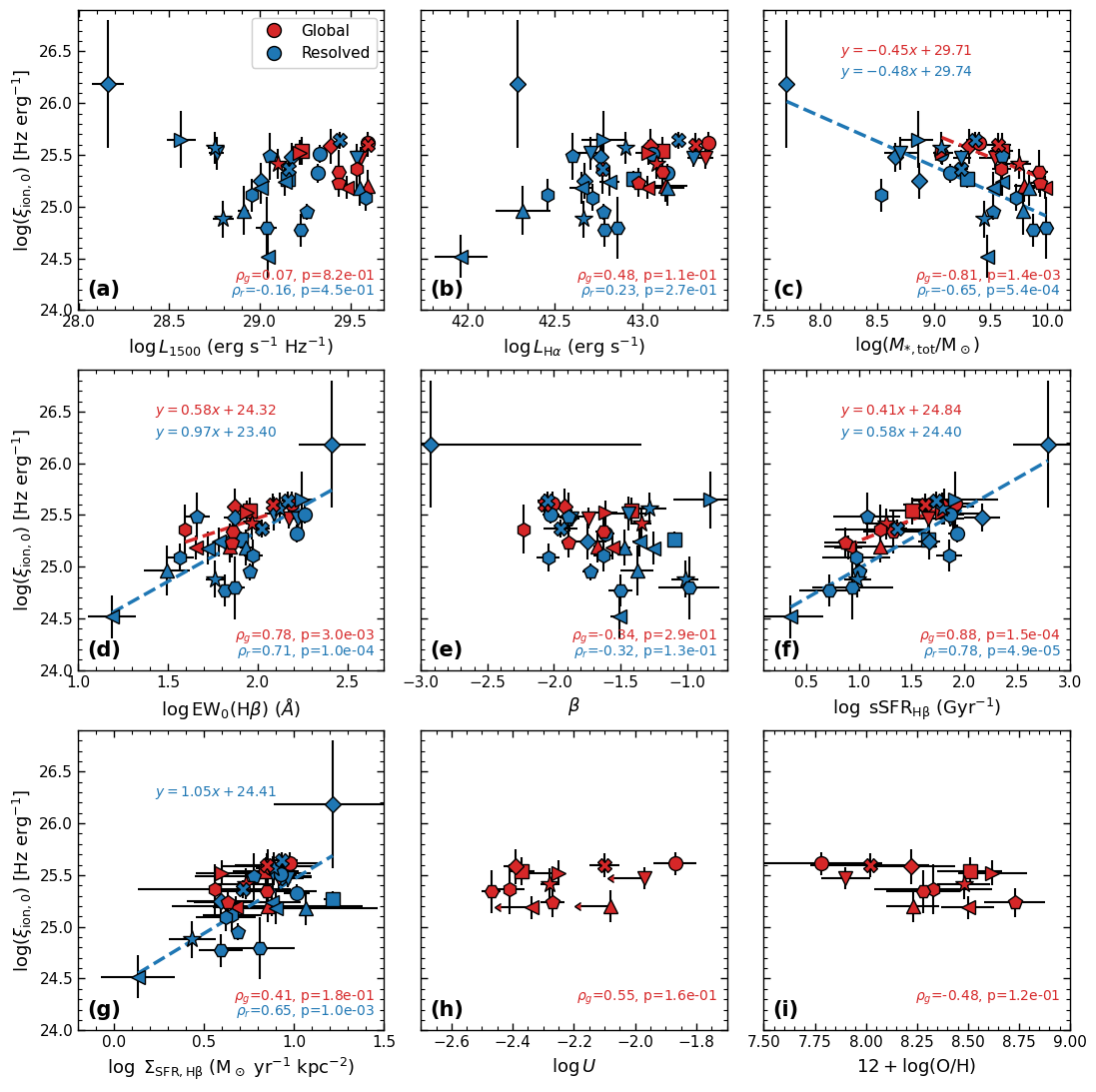} 
\caption{Comparison of \logxio\ in the REBELS-IFU sample to physical properties on the galaxy (red markers) and individual-clump (blue markers) scale: a) observed rest-frame $1500$~\AA~luminosity density; b) observed \Ha\ luminosity; c) $M_{*,\rm tot}$; d) $\rm EW_0(H\beta)$; e) UV slope $\beta$; f) $\rm sSFR_{H\beta}$; g) $\Sigma_{\mathrm{SFR, H\beta}}$; h) $\log(U)$, \textbf{with arrows representing upper limits in $z>7$ galaxies without Balmer decrement measurements}; and i) $12+\log(\rm O/H)$. Spearman's rank correlation coefficients $\rho$ and p-values are shown for each subset of measurements in the corresponding colors in the bottom-right corner.  For the significant correlations, we show the linear regression results with dashed lines and linear equations of the corresponding color. The same marker convention is used as in Figure~\ref{fig:global_resolved}. }
\label{fig:Xicorr}
\end{figure*}

The relations of \Xio\ and the above properties are shown in Figure~\ref{fig:Xicorr}. We quantify the degree of correlation between \Xio\ and the above parameters using the Spearman's rank correlation test. We consider correlations significant if the Spearman’s coefficient $|\rho| \geq 0.4$, and the p-value is $p \leq 0.05$. The resulting correlation coefficients $\rho$ and p-values for each pair of variables are listed in Table~\ref{table:corr}. For the significant correlations, we present linear regression results in the corresponding panels of Figure~\ref{fig:Xicorr}. 

In the galaxy-scale \Xio\ measurements, we find statistically significant correlations with EW(H$\beta$) ($\rho=0.78, p=3.0\times10^{-3}$) and specific star formation rate $\rm sSFR_{\rm H\beta}$ ($\rho=0.88, p=1.5\times10^{-4}$), and an anti-correlation with total stellar mass ($\rho=-0.81, p=1.4\times10^{-3}$). So, higher global ionizing photon production efficiencies are found in younger, more star-forming, less massive galaxies, as expected from previous studies \citep[e.g.,][]{Castellano2023, Llerena2024_Xi, Jaiswar2024}.

Quantifying these galaxy-scale relations, our linear regression analysis yields \logxio\ = $(0.58\pm 0.21) \times \log \rm EW(H\beta) + (24.32\pm 0.40)$; \logxio\ = $(0.41\pm 0.06) \times \log \rm sSFR_{\rm H\beta} + (24.84\pm 0.09)$; \logxio\ = $(-0.45 \pm 0.11) \times \log M_{*,\rm tot}/M_\odot + (29.71 \pm 1.08)$. The slope is consistent with previously reported values of $\sim -0.4$ for $M_{*,\rm tot}$, and $\sim 0.4$ to $1.2$ for EW(H$\beta$), but steeper than $\sim 0.2$ found for sSFR \citep{Chevallard2018, Tang2019, Llerena2024_Xi, Castellano2023, PrietoLyon2023}. Although notably, the latter was reported for a significantly larger sample, spanning $\geq 3$~dex in mass, and is not directly comparable. 
Overall, our results show that even in samples as small as ours ($N=12$), EW(H$\beta$), $M_{*,\rm tot}$, and sSFR are significant predictors of \Xio. 

With the sample of 25 individual star-forming clumps within the REBELS-IFU galaxies, we can probe the link between ionizing photon production and local physical conditions on the scale of a couple kpc. Although as noted in Section~\ref{sec:spatialbinning}, the physical nature of these clumps is ambiguous, as the individual nebular morphological components may trace separate merging galaxies, or large H~II complexes within one galaxy.

On the clump scale, we recover the same three correlations seen on the galaxy scale, along with one additional correlation. First, as on the global galaxy scale, we find that the local efficiency \Xio\ increases with the EW(H$\beta$) ($\rho=0.71, p=1.0\times10^{-4}$) and specific star formation rate $\rm sSFR_{\rm H\beta}$ ($\rho=0.71, p=2.3\times10^{-4}$), and decreases with clump stellar mass ($\rho=-0.58, p=3.0\times10^{-3}$). Compared to the galaxy-scale relations, these clump-scale correlations span additional $0.5-1.5$ dex. Thus, more dense, intensely star-forming clumps produce more ionizing photons per unit rest-UV luminosity. The strong correlations of \Xio\ with sSFR and EW(H$\beta$) are particularly expected, given that \Xio\ decreases with increasing stellar age and metallicity \citep{Raiter2010, Stanway2020}, which are in turn anti-correlated with EW(H$\beta$) and sSFR \citep{Mannucci2010, Gozaliasl2024}. Second, the clump \Xio\ correlates with the star formation rate surface density $\Sigma_{\mathrm{SFR, H\beta}}$ ($\rho=0.65, p=1.0\times10^{-3}$). The lack of a significant correlation on the global scale suggests that smaller-scale star-forming conditions drive the ionizing photon production efficiency.
So, younger, denser, less massive star-forming clumps have increasingly higher ionizing photon production efficiencies. 

\begin{deluxetable*}{lccccccc}
\tablecaption{REBELS-IFU Galaxy Properties
\label{table:globalXi}}
\tablehead{
\colhead{ID} & \colhead{$L_{{\mathrm{H}\alpha}}$ \tablenotemark{\rm \scriptsize a}} & \colhead{$\Sigma_{\mathrm{SFR, H\beta}}$ \tablenotemark{\rm \scriptsize b}} & \colhead{$\rm sSFR_{H\beta}$ \tablenotemark{\rm \scriptsize c}} & \colhead{$\mathrm{EW_0}(\mathrm{H}\beta)$ \tablenotemark{\rm \scriptsize d}} & \colhead{$A_{\rm V, neb}$ \tablenotemark{\rm \scriptsize e}} & \colhead{$A_{\rm V, UV}$ \tablenotemark{\rm \scriptsize f}} & \colhead{$\log(\xi_{\rm ion,0})$ \tablenotemark{\rm \scriptsize g}}}
\startdata
REBELS-05 & $1.30 \pm 0.04$ & $7.0 \pm 3.0$ & $32 \pm 15$ & $90 \pm 3$ & $0.81 \pm 0.37$ & $0.37 \pm 0.09$ & $25.54 \pm 0.13$ \\
REBELS-08 & $1.12 \pm 0.04$ & $7.1 \pm 4.2$ & $65 \pm 41$ & $74 \pm 3$ & $1.09 \pm 0.48$ & $0.28 \pm 0.06$ & $25.58 \pm 0.17$ \\
REBELS-12 & $1.36 \pm 0.39^\dagger$ & $7.1 \pm 8.7$ & $16 \pm 20$ & $70 \pm 3$ & $0.45 \pm 0.29^\dagger$ & $0.18 \pm 0.09$ & $25.20 \pm 0.15$ \\
REBELS-14 & $2.29 \pm 0.24^\dagger$ & $8.2 \pm 2.2$ & $46 \pm 18$ & $149 \pm 4$ & $0.36 \pm 0.21^\dagger$ & $0.17 \pm 0.06$ & $25.47 \pm 0.10$ \\
REBELS-15 & $2.38 \pm 0.06$ & $9.5 \pm 3.4$ & $82 \pm 30$ & $154 \pm 4$ & $0.64 \pm 0.32$ & $0.34 \pm 0.06$ & $25.61 \pm 0.11$ \\
REBELS-18 & $1.07 \pm 0.14^\dagger$ & $4.7 \pm 1.3$ & $8 \pm 2$ & $45 \pm 1$ & $0.40 \pm 0.22^\dagger$ & $0.26 \pm 0.05$ & $25.19 \pm 0.11$ \\
REBELS-25 & $1.08 \pm 0.18^\dagger$ & $4.0 \pm 1.3$ & $62 \pm 25$ & $86 \pm 6$ & $0.29 \pm 0.22^\dagger$ & $0.25 \pm 0.06$ & $25.52 \pm 0.12$ \\
REBELS-29 & $0.95 \pm 0.03$ & $4.3 \pm 1.9$ & $7 \pm 3$ & $71 \pm 3$ & $0.31 \pm 0.38$ & $0.33 \pm 0.08$ & $25.23 \pm 0.14$ \\
REBELS-32 & $1.21 \pm 0.03$ & $5.4 \pm 2.0$ & $18 \pm 8$ & $93 \pm 3$ & $0.62 \pm 0.33$ & $0.45 \pm 0.12$ & $25.41 \pm 0.15$ \\
REBELS-34 & $0.60 \pm 0.03$ & $3.6 \pm 3.5$ & $16 \pm 16$ & $39 \pm 3$ & $0.94 \pm 0.72$ & $0.10 \pm 0.04$ & $25.36 \pm 0.24$ \\
REBELS-38 & $1.31 \pm 0.04$ & $7.1 \pm 4.5$ & $21 \pm 14$ & $72 \pm 3$ & $1.25 \pm 0.54$ & $0.58 \pm 0.14$ & $25.34 \pm 0.21$ \\
REBELS-39 & $2.02 \pm 0.06$ & $7.0 \pm 2.8$ & $42 \pm 20$ & $120 \pm 3$ & $0.52 \pm 0.34$ & $0.17 \pm 0.04$ & $25.60 \pm 0.11$ \\
\enddata
\textbf{Notes.} 
\tablenotetext{}{\textbf{Global line fluxes are adopted from \cite{Rowland2025}.}}
\tablenotetext{\rm \dagger}{\Ha\ is outside of the NIRSpec coverage, and \Ha\ luminosity and nebular attenuation are estimated from \Hb\ and the UV continuum attenuation (Section~\ref{sec:analysis}).}
\tablenotetext{\rm a}{Observed \Ha\ luminosity in $10^{43} \rm~erg~s^{-1}$.}
\tablenotetext{\rm b}{Star formation rate surface density in $\rm M_\odot \rm~yr^{-1}~kpc^{-2}$, measured from the de-reddened \Hb\ luminosity and the area of the spectral extraction masks (Stefanon et al. in prep).}
\tablenotetext{\rm c}{Specific star formation rate in $\rm Gyr^{-1}$, obtained from de-reddened \Hb\ luminosity and stellar masses derived by \cite{Fisher2025}. }
\tablenotetext{\rm d}{Rest-frame equivalent width of \Hb\ in $\mathrm{\AA}$, which we measure with our best spectral fits to the line and continuum (Section~\ref{sec:analysis}). }
\tablenotetext{\rm e}{V-band dust attenuation of the nebular gas, computed from the Balmer decrement and custom attenuation curves derived by \cite{Fisher2025}. These differ slightly from those presented by \cite{Rowland2024} because they adopt the \cite{Calzetti2000} law.}
\tablenotetext{\rm f}{V-band dust attenuation of the stellar continuum, obtained with SED fitting by \cite{Fisher2025}.}
\tablenotetext{\rm g}{$\log10$ of the global ionizing photon production efficiency in $\rm Hz~erg^{-1}$.}
\end{deluxetable*}

\movetabledown=2.3in
\begin{rotatetable*} 
\begin{deluxetable*}{lccccccccccc}
\tablecaption{Physical Parameters of Individual Clumps in REBELS-IFU Galaxies
\label{table:resolvedXi}}
\tablehead{
\colhead{ID} & \colhead{$L_{{\mathrm{H}\alpha}}$} & \colhead{$L_{\mathrm{1500}}$ } & \colhead{$\log(M_{*,\rm tot}/\rm M_{\odot})$} & \colhead{$\beta$} & \colhead{$\rm SFR_{\mathrm{H\beta}}$}&  \colhead{$\rm sSFR_{\mathrm{H\beta}}$} & \colhead{$\Sigma_{\mathrm{SFR, H\beta}}$ \tablenotemark{\rm \scriptsize a}}& \colhead{$\mathrm{EW}(\mathrm{H}\beta)$} & \colhead{$A_{\rm V, neb}$ \tablenotemark{\rm \scriptsize b}} & \colhead{$A_{\rm V, UV}$ \tablenotemark{\rm \scriptsize c}} & \colhead{$\log(\xi_{\rm ion,0})$} \\
\colhead{} & \colhead{ $10^{42} \rm~erg~s^{-1}$} & \colhead{$\rm 10^{28}~erg~s^{-1}~Hz^{-1}$} & \colhead{} & \colhead{} & \colhead{$\rm M_\odot \rm~yr^{-1}$} & \colhead{$\rm Gyr^{-1}$} & \colhead{$\rm M_\odot \rm~yr^{-1}~kpc^{-2}$} & \colhead{\AA} & \colhead{} & \colhead{} & \colhead{}}
\startdata
R-05-A & $8.9 \pm 0.1$ & $14.3 \pm 0.5$ & $9.30_{-0.05}^{+0.07}$ & $-1.09 \pm 0.06$ & $91 \pm 10$ & $46 \pm 7$ & $16 \pm 2$ & $80 \pm 3$ & $0.9 \pm 0.1$ & $0.5 \pm 0.1$ & $25.27 \pm 0.08$ \\
\hline
R-08-A & $5.8 \pm 0.2$ & $15.1 \pm 0.5$ & $8.66_{-0.01}^{+0.03}$ & $-1.89 \pm 0.08$ & $68 \pm 26$ & $148 \pm 59$ & $9 \pm 3$ & $74 \pm 10$ & $1.1 \pm 0.3$ & $0.3 \pm 0.1$ & $25.48 \pm 0.14$ \\
R-08-B & $4.7 \pm 0.2$ & $10.1 \pm 0.6$ & $8.87_{-0.01}^{+0.02}$ & $-1.75 \pm 0.12$ & $34 \pm 15$ & $47 \pm 11$ & $4 \pm 2$ & $70 \pm 9$ & $0.5 \pm 0.3$ & $0.4 \pm 0.1$ & $25.25 \pm 0.17$ \\
R-08-C & $1.9 \pm 0.1$ & $1.4 \pm 0.3$ & $7.70_{-0.03}^{+0.04}$ & $-2.92 \pm 1.58$ & $31 \pm 24$ & $619 \pm 469$ & $16 \pm 12$ & $258 \pm 110$ & $1.5 \pm 0.6$ & -- & $26.19 \pm 0.61$ \\
\hline
R-12-A & $2.1 \pm 0.7^\dagger$ & $8.1 \pm 0.6$ & $9.79_{-0.01}^{+0.01}$ & $-1.38 \pm 0.12$ & -- & -- & -- & $31 \pm 9$ & $0.7 \pm 0.5^\dagger$ & $0.3 \pm 0.1$ & $24.96 \pm 0.24$ \\
R-12-B & $13.8 \pm 3.1^\dagger$ & $35.4 \pm 0.8$ & $9.84_{-0.06}^{+0.05}$ & $-1.47 \pm 0.04$ & $115 \pm 107$ & -- & $12 \pm 11$ & $86 \pm 7$ & $0.6 \pm 0.4^\dagger$ & $0.3 \pm 0.1$ & $25.18 \pm 0.17$ \\
\hline
R-14-A & $19.6 \pm 1.6^\dagger$ & $34.4 \pm 0.8$ & $9.24_{-0.07}^{+0.08}$ & $-1.86 \pm 0.06$ & $129 \pm 27$ & $74 \pm 21$ & $9 \pm 2$ & $167 \pm 11$ & $0.3 \pm 0.2^\dagger$ & $0.1 \pm 0.1$ & $25.47 \pm 0.08$ \\
R-14-B & $5.0 \pm 0.8^\dagger$ & $5.8 \pm 0.6$ & $8.70_{-0.14}^{+0.15}$ & $-1.44 \pm 0.15$ & $40 \pm 19$ & $79 \pm 49$ & $8 \pm 4$ & $123 \pm 15$ & $0.5 \pm 0.3^\dagger$ & $0.3 \pm 0.1$ & $25.52 \pm 0.16$ \\
\hline
R-15-A & $13.8 \pm 0.2$ & $21.0 \pm 0.9$ & $9.14_{-0.01}^{+0.01}$ & $-1.61 \pm 0.07$ & $118 \pm 19$ & $85 \pm 15$ & $10 \pm 2$ & $165 \pm 10$ & $0.6 \pm 0.2$ & $0.5 \pm 0.1$ & $25.32 \pm 0.09$ \\
R-15-B & $11.5 \pm 0.2$ & $21.4 \pm 1.0$ & $9.07_{-0.01}^{+0.01}$ & $-2.03 \pm 0.06$ & $84 \pm 18$ & $73 \pm 22$ & $8 \pm 2$ & $181 \pm 15$ & $0.4 \pm 0.2$ & $0.2 \pm 0.1$ & $25.51 \pm 0.09$ \\
\hline
R-18-A & $6.5 \pm 0.9^\dagger$ & $13.7 \pm 0.3$ & $9.60_{-0.02}^{+0.02}$ & $-1.36 \pm 0.05$ & $50 \pm 18$ & -- & $8 \pm 3$ & $60 \pm 3$ & $0.5 \pm 0.3^\dagger$ & $0.3 \pm 0.1$ & $25.24 \pm 0.14$ \\
R-18-B & $4.5 \pm 0.7^\dagger$ & $10.1 \pm 0.3$ & $9.52_{-0.04}^{+0.04}$ & $-1.25 \pm 0.05$ & $36 \pm 15$ & -- & $8 \pm 3$ & $53 \pm 4$ & $0.6 \pm 0.3^\dagger$ & $0.3 \pm 0.1$ & $25.18 \pm 0.16$ \\
R-18-C & $0.9 \pm 0.3^\dagger$ & $11.1 \pm 0.5$ & $9.47_{-0.04}^{+0.04}$ & $-1.51 \pm 0.06$ & $7 \pm 3$ & $2 \pm 2$ & $1 \pm 1$ & $15 \pm 5$ & $0.4 \pm 0.2^\dagger$ & $0.2 \pm 0.1$ & $24.52 \pm 0.21$ \\
\hline
R-25-A & $6.0 \pm 1.7^\dagger$ & $3.7 \pm 0.7$ & $8.86_{-0.15}^{+0.13}$ & $-0.82 \pm 0.28$ & -- & $81 \pm 77$ & -- & $174 \pm 26$ & $0.8 \pm 0.6^\dagger$ & $0.4 \pm 0.1$ & $25.65 \pm 0.27$ \\
\hline
R-29-A & $6.0 \pm 0.1$ & $18.2 \pm 0.5$ & $9.52_{-0.04}^{+0.04}$ & $-1.72 \pm 0.06$ & $34 \pm 1$ & $10 \pm 2$ & $5 \pm 0$ & $90 \pm 6$ & -- & $0.3 \pm 0.1$ & $24.95 \pm 0.08$ \\
R-29-B & $4.0 \pm 0.3$ & $11.4 \pm 0.6$ & $9.60_{-0.05}^{+0.06}$ & $-1.89 \pm 0.13$ & $48 \pm 36$ & $12 \pm 9$ & $6 \pm 4$ & $46 \pm 8$ & $1.1 \pm 0.6$ & $0.2 \pm 0.1$ & $25.49 \pm 0.23$ \\
\hline
R-32-A & $8.0 \pm 0.2$ & $5.6 \pm 0.5$ & $9.06_{-0.03}^{+0.02}$ & $-1.28 \pm 0.13$ & $71 \pm 17$ & $61 \pm 5$ & $8 \pm 2$ & $132 \pm 11$ & $0.7 \pm 0.2$ & $0.6 \pm 0.1$ & $25.57 \pm 0.15$ \\
R-32-B & $4.6 \pm 0.2$ & $6.3 \pm 0.8$ & $9.44_{-0.03}^{+0.04}$ & $-1.02 \pm 0.10$ & $27 \pm 8$ & $10 \pm 3$ & $3 \pm 1$ & $58 \pm 7$ & -- & $0.8 \pm 0.1$ & $24.88 \pm 0.18$ \\
\hline
R-34-A & $2.9 \pm 0.1$ & $9.1 \pm 0.3$ & $8.54_{-0.01}^{+0.02}$ & $-1.63 \pm 0.08$ & $25 \pm 9$ & $71 \pm 20$ & $5 \pm 2$ & $94 \pm 11$ & $0.6 \pm 0.3$ & $0.4 \pm 0.1$ & $25.11 \pm 0.16$ \\
R-34-B & $5.2 \pm 0.2$ & $38.3 \pm 1.6$ & $9.72_{-0.05}^{+0.04}$ & $-2.04 \pm 0.08$ & $50 \pm 19$ & $9 \pm 7$ & $4 \pm 2$ & $37 \pm 7$ & $0.8 \pm 0.3$ & $0.1 \pm 0.1$ & $25.09 \pm 0.13$ \\
\hline
R-38-A & $6.1 \pm 0.2$ & $16.9 \pm 0.6$ & $9.88_{-0.03}^{+0.03}$ & $-1.50 \pm 0.09$ & $40 \pm 11$ & $5 \pm 4$ & $4 \pm 1$ & $65 \pm 5$ & $0.3 \pm 0.2$ & $0.8 \pm 0.1$ & $24.77 \pm 0.16$ \\
R-38-B & $7.2 \pm 0.4$ & $10.9 \pm 1.4$ & $9.99_{-0.09}^{+0.07}$ & $-0.99 \pm 0.23$ & $85 \pm 38$ & $9 \pm 8$ & $6 \pm 3$ & $75 \pm 10$ & $1.0 \pm 0.4$ & $1.4 \pm 0.3$ & $24.80 \pm 0.30$ \\
\hline
R-39-A & $16.1 \pm 0.6$ & $27.6 \pm 1.2$ & $9.36_{-0.08}^{+0.07}$ & $-2.05 \pm 0.07$ & $124 \pm 26$ & $54 \pm 27$ & $9 \pm 2$ & $146 \pm 11$ & $0.5 \pm 0.2$ & $0.1 \pm 0.1$ & $25.64 \pm 0.08$ \\
R-39-B & $5.9 \pm 0.1$ & $14.5 \pm 0.6$ & $9.24_{-0.07}^{+0.06}$ & $-1.95 \pm 0.13$ & $40 \pm 1$ & $23 \pm 5$ & $5 \pm 0$ & $105 \pm 10$ & $0.3 \pm 0.1$ & $0.2 \pm 0.1$ & $25.37 \pm 0.09$ \\
\hline
\enddata
\tablecomments{}
\tablenotetext{\rm \dagger}{\Ha\ is outside of the NIRSpec coverage, and \Ha\ luminosity and nebular attenuation are obtained from \Hb\ and UV slope $\beta$.}
\tablenotetext{\rm a}{Star formation rate surface density, measured from the de-reddened \Hb\ luminosity and the pixel area of each clump, obtained from clump extraction masks (Section~\ref{sec:spatialbinning}).}
\tablenotetext{\rm b}{V-band dust attenuation of the nebular gas component, which we derive from the Balmer decrement.}
\tablenotetext{\rm c}{V-band dust attenuation of the stellar continuum, which we derive from the UV slope $\beta$, assuming the \cite{Fisher2025} extinction laws.}
\end{deluxetable*}
\end{rotatetable*}

The quantitative relations established by our linear regression analysis of local \Xio\ vs. $M_{*,\rm tot}$ is consistent with the global one within uncertainties. However, the relation with local $\rm EW_0(H\beta)$ is significantly steeper,  \logxio\ = $(0.97\pm 0.18) \times \log \rm EW(H\beta) + (23.40\pm 0.34)$. This implies a strong physical coupling between \Xio\ and the youth of the stellar population, while the shallower global relation reflects spatial averaging that dilutes this intrinsic trend. The same reasoning applies to the local correlation of \Xio\ with sSFR, where the slope is somewhat steeper on the clump scale, $0.58\pm0.07$ vs. $0.41\pm0.06$ globally. Finally, the additional local correlation with $\Sigma_{\mathrm{SFR, H\beta}}$ is parameterized as \logxio\ = $(1.05\pm 0.22) \times \log \Sigma_{\mathrm{SFR, H\beta}} + (24.41\pm 0.19)$.

The parameters that do not show significant correlations with \Xio\ are also informative. First, we do not recover a correlation of \Xio\ with redshift that has been presented in the literature \citep[e.g.,][]{Matthee2017}. This may be expected, however, given the relatively narrower redshift range of our sample, $z = 6.50-7.67$. \textbf{Second, REBELS galaxies show substantial dust content \citep{Inami2022, Ferrara2022, Sommovigo2022, Algera2025, Fisher2025b}, and it is important to test whether dustiness is linked to lower \Xio\ in this sample. We do not find any correlations between \Xio\ and observed UV slope $\beta$, $A_{\rm V, neb}$, or $A_{\rm V, UV}$. This may be expected, because \Xio\ is fundamentally set by intrinsic stellar population properties, and dust content can only be linked to \Xio\ indirectly, e.g. via metallicity. It is instead likely more important to ionizing photon escape \citep[e.g.,][]{Gazagnes2020, Ji2025}.} Third, given the lack of correlations with the observed \Ha\,, [C~II], and rest-UV luminosities, we see that higher \Xio\ is not just a matter of brighter emission lines or fainter UV luminosities. This shows that the correlations of \Xio\ with sSFR, $\Sigma_{\mathrm{SFR, H\beta}}$, and $\rm EW_0(H\beta)$ are not simply driven by stronger recombination lines, but indeed demonstrate the physical origin of higher ionizing photon production efficiencies. Finally, we note the lack of correlation of \Xio\ with $12+\log(\rm O/H)$, \textbf{derived from $R_{23}$,} and \textbf{$\log(U)$, derived from $O_{32}$, by \cite{Rowland2025}}. This shows that in the REBELS-IFU sample, \Xio\ is not significantly linked to the ionization state and metallicity of the nebular gas. \textbf{This is not consistent with previous studies, which have found \Xio\ to correlate with $O_{32}$ and $R_{23}$ \citep{Pahl2025, Llerena2024_Xi}. However, these correlations, based on significantly larger samples, show a large scatter of $0.5-1.0$~dex. It is thus likely that we do not sufficiently sample the parameter space to recover them. } 

\begin{deluxetable}{lcccc}
\tablecaption{\Xio\ vs. Galaxy and Clump Properties: Spearman's Rank Correlation Results 
\label{table:corr}}
\tablehead{
\colhead{Parameter} & \multicolumn{2}{c}{$\xi_{\rm ion, 0, global}$} & \multicolumn{2}{c}{$\xi_{\rm ion, 0, clump}$}\\
\colhead{} & \colhead{$\rho$} & \colhead{$p$} & \colhead{$\rho$} & \colhead{$p$}}
\startdata
$z$ & -0.16 & 0.62 & 0.24 & 0.25\\
$L_{1500}$ & 0.07 & 0.82 &  -0.16 & 0.45 \\
$L_{\rm H\alpha}$ & 0.48 & 0.11 & 0.23 & 0.27 \\
$L_{\rm [C~II] 158\mu m}$ & -0.28 & 0.43 & -- & -- \\
$\beta$  & -0.33 & 0.29 & -0.32 & 0.13 \\
$\rm EW_0(H\beta)$ & \textbf{0.78} & \boldmath{$3.0\times10^{-3}$}  & \textbf{0.71} & \boldmath{$1.0\times10^{-4}$} \\
$\rm sSFR_{H\beta}$ & \textbf{0.88} & \boldmath{$1.5\times10^{-4}$} & \textbf{0.78} & \boldmath{$4.9\times10^{-5}$} \\
$\Sigma_{\mathrm{SFR, H\beta}}$ & {0.41} & {0.18} & \textbf{0.65} & \boldmath{$1.0\times10^{-3}$} \\
$\log(U)$ & 0.55 & 0.16 & -- & -- \\
$12 + \log(\rm O/H)$ & -0.47 & 0.12 & -- & -- \\
$M_{*,\rm tot}$ & \textbf{-0.81} & \boldmath{$1.3\times10^{-3}$} & \textbf{-0.65} & \boldmath{$5.4\times10^{-4}$} \\
\enddata 
\textbf{Notes.} \\
We boldface the significant correlations with $|\rho| \geq 0.4$, $p \leq 0.05$.
\end{deluxetable}

\begin{figure*}
\includegraphics[width=0.9\textwidth]{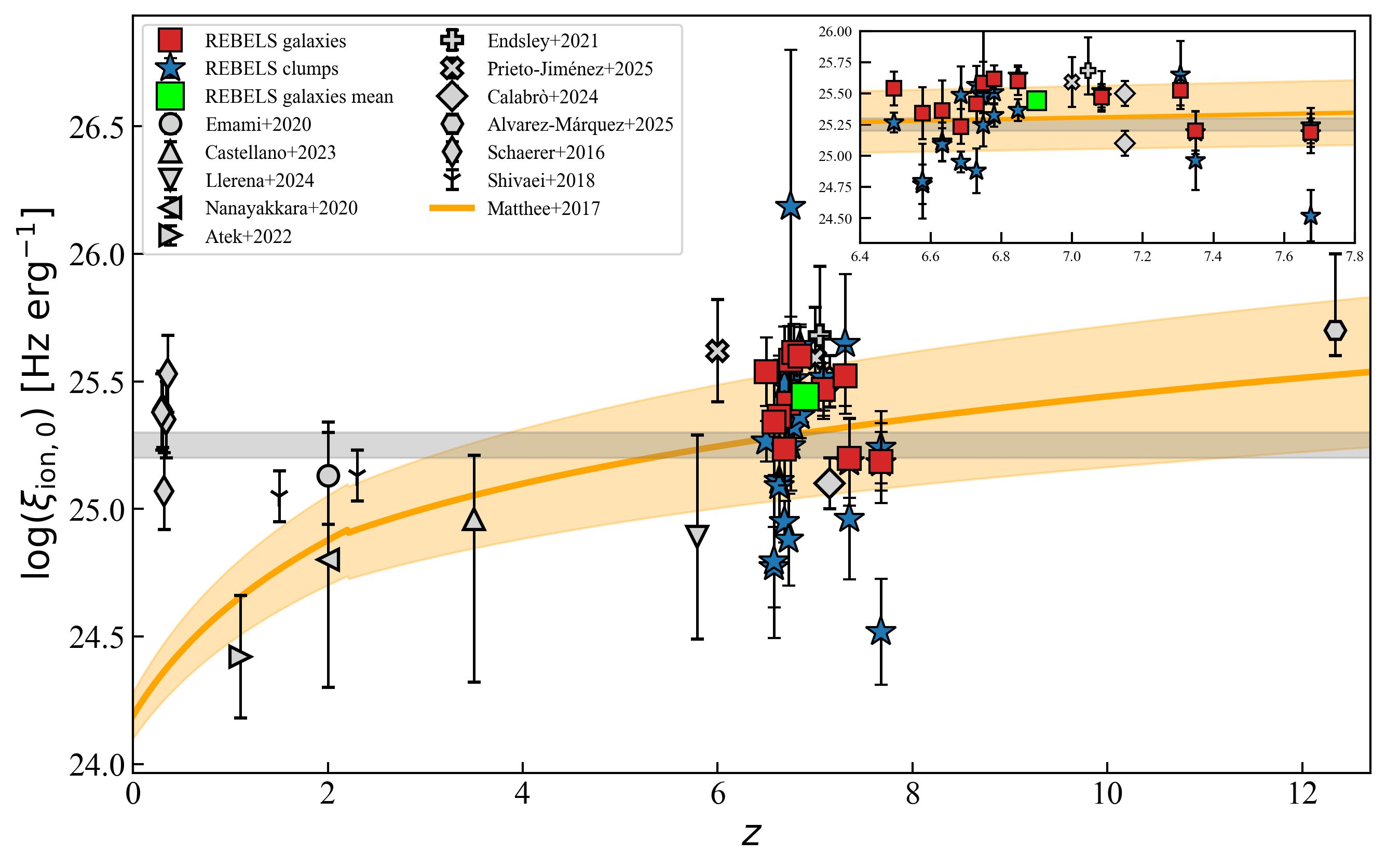} 
\caption{Comparison of \logxio\ in the REBELS-IFU sample to measurements in the literature across redshift, \textbf{for $M_* \geq 10^{8.7}~\rm M_{\odot}$}, with various studies represented by different grey symbols, as indicated in the legend. Galaxy-scale REBELS measurements are shown as red squares, with the green square representing the weighted mean of the sample. Resolved, clump-scale measurements are shown as blue stars. The canonical value of \Xio\ is shown as a grey band. The yellow curve is the parameterization of the redshift evolution in \Xio\ reported by \cite{Matthee2017} \textbf{for $\log(M_*/\rm M_{\odot}) = 9.2$. We adopt this curve because, although it lies at the lower bound of the REBELS mass range, it is closer to the median stellar mass of the massive galaxy samples compiled from the literature.} \textbf{The inset in the top-right shows a zoom-in of the REBELS-IFU redshift range.}}
\label{fig:litcomp}
\end{figure*}

\section{Discussion}
\label{sec:discussion}
We have presented the first spatially resolved measurements of the ionizing photon production efficiency \Xio\ in a sample of exceptionally bright ($M_\mathrm{UV} \sim -22$~mag) EoR galaxies. Our spatially resolved data reveal that global measurements dilute a much larger range of smaller-scale \Xio\ values, ranging from $0.2\times$ to $4\times$ the corresponding global values. These variations in \Xio\ are linked to local variations in stellar population age and star formation history, traced by EW(H$\beta$) and sSFR, stellar mass, and star formation rate surface density. In this section, we discuss how our findings compare to those of previous studies in the literature, as well as the implications of these results to our understanding of reionization.
 
In Figure~\ref{fig:litcomp}, we compare our measured values of \Xio\ to a compilation of \Xio\ across cosmic time reported in the literature \citep{Schaerer2016, Matthee2017, Shivaei2018, Emami2020,Nanayakkara2020,  Endsley2021, Atek2022, Castellano2023, Llerena2024_Xi, Calabro2024, PrietoJimenez2025, AlvarezMarquez2025}. \textbf{Specifically, we select existing \Xio\ measurements in galaxies with high stellar masses $M_* \geq 10^{8.7}~\rm M_{\odot}$, such that they are comparable to the REBELS-IFU galaxy masses of $10^{9.2} - 10^{9.7}~\rm M_{\odot}$. Additionally, we show this comparison for all available stellar masses in the appendix in Figure~\ref{fig:app_lit}.} As can be seen in Figure~\ref{fig:litcomp}, our global \Xio\ measurements are largely consistent with those previously reported for \textbf{massive galaxies in} the redshift range $z = 6-8$, spanning \logxio = $25.5-26.0$ \citep{Stark2015, Stark2017, Endsley2021, Stefanon2022, Matthee2023,  Llerena2024_Xi,Heintz2025}. The efficiencies of REBELS-IFU galaxies extend to somewhat lower values, reaching \logxio = $25.19\pm0.11$, consistent with the one canonically assumed in reionization models. \textbf{A likely explanation for this is that the REBELS galaxies are massive and reach lower sSFRs \citep{Topping2022}, and \Xio\ correlates with sSFR \citep{Llerena2024_Xi, Castellano2023}. The significant dust content of REBELS galaxies is unlikely to drive the lower \Xio\, as our measurements are corrected for dust, and reflect intrinsic stellar population properties.} This also explains the finding that REBELS galaxies fall on the low-excitation and -ionization tail of typical star-forming galaxies at $z \sim 7$ \citep{Endsley2025}.
Our measured global \Xio\ values are also consistent with the redshift evolution of \Xio\ parameterized by \cite{Matthee2017}, falling within $1\sigma$ of the best fit $\xi_{\rm ion,0}(z)$. The mean galaxy-integrated \Xio\ of our sample $25.44\pm0.15$  is indeed consistent with the predicted \logxio\ $=25.29\pm0.25$ at our mean redshift of $z = 6.9$. 

The resolved, clump-scale measurements of \Xio\ show a much larger scatter compared to the sample of galaxy-scale measurements in the literature, spanning almost two orders of magnitude \logxio = $24.51-26.18$. As a result, the \Xio\ values in individual star-forming regions significantly deviate from the galaxy-averaged \Xio\ redshift evolution. This local diversity is consistent with the highest-redshift spatially resolved measurements of \Xio\, in the lensed galaxy MACS1149-JD1 at  $z =9.1$ \citep{AlvarezMarquez2024}, where the global value is \logxio=$25.5 \pm0.03$, and the resolved clump N reaches $25.9 \pm0.09$. 

\subsection{The Nature of Star-Forming Clumps in Massive EoR Galaxies}
It is important to discuss the possible physical nature of what we have been referring to as ``clumps'' in this work. In our spatially resolved analysis of the REBELS-IFU sample, we have decomposed the nebular (H$\alpha$ or H$\beta$) images of each galaxy into $1-3$ components. A first look at the recently obtained NIRCam observations for 7/12 of our galaxies (PID 6480, PI: Schouws) shows these structures are indeed recovered in the higher-resolution F115W, F200W, F356W imaging, with some galaxies appearing to have even more clumps than resolved with NIRSpec IFU. These images and their detailed analysis will be presented in upcoming works from the REBELS collaboration. 

Our clumps have circularized radii of $0.78-2.15$~kpc and stellar massses in the range $\log(M_{*,\rm tot}/\rm M_{\odot}) = {7.70}-{9.71}$. Typical super star clusters have masses up to $10^6 -10^7 \rm~M_{\odot}$ and radii $1-10$~pc \citep{Portegies2010, Brown2021}, although they can produce H~II regions up to kpc scales \citep[e.g.,][]{TenorioTagle2006}. Our clumps are nevertheless too massive to be individual clusters, and are instead either lower-mass galaxies in the process of merging, or large H~II complexes within the same galaxy. For the latter possibility, they may be similar to star-forming knots seen in local, spatially resolved irregular starbursts. For example, Haro~11 at $z = 0.02$ is comprised of three kpc-scale H~II complexes, each containing multiple clusters with total masses of $10^7 - 10^8~\rm M_{\odot}$ \citep{Sirressi2022}. From our measurements so far, we see that properties such as $M_{*,\rm tot}$, $\beta$, $\rm SFR_{H\beta}$, EW(H$\beta$), and \Xio\ are significantly divergent among the clumps in most of our galaxies (Table~\ref{table:resolvedXi}). However, to conclusively distinguish between the in-situ vs. ex-situ scenarios for the origin of the clumps, further investigation into the metallicities and kinematics of the clumps is required. 

The clumpy, irregular morphologies of the REBELS galaxies are qualitatively consistent with previous observations of high-redshift galaxies. For example, a number of JADES galaxies at $z \geq 8$ show multiple UV clumps accompanied by diffuse nebular emission \citep{Hainline2024}. \cite{Elmegreen2005, Elmegreen2007} identify highly clumpy galaxies at $z = 1-5$ in the Hubble Ultra Deep Field, with clumps of radii $\sim 0.5-1$~kpc and masses $10^8-10^9 \rm~M_{\odot}$. FirstLight simulations find that most of such clumps at $z \sim 5-9$ are formed through mergers \citep{Nakazato2024}. If these clumps have sufficiently low star formation efficiencies and survive under radiation pressure, they will likely migrate into the center within several dynamic times \citep{Krumholz2010, Mandelker2017}. They would therefore represent early stages of disk formation, where the centrally migrated clumps eventually form the bulge.

\subsection{Extreme Values of $\xi_{\rm ion,0}$}

Our spatially resolved measurements reveal one possibly extreme clump-scale values of \logxio\ in REBELS-08-C, reaching $26.19\pm0.61$. This clump, as can be seen in Figure~\ref{fig:2dfits_app}, is spatially offset from the rest of the galaxy, and may in fact be a lower-mass merging component. Its high \Xio\ value, though with a large uncertainty, is not only significantly above the canonical $25.3$, but is at the upper limit of the expected range of \Xio\ for normal stellar populations. Traditional stellar population models with, e.g. a Chabrier IMF, predict a maximum intrinsic \logxio $=25.8-26.1$, \textbf{for the youngest stellar ages} \citep{Chen2024, Katz2024}. \textbf{It is notable that the high \Xio\ in REBELS-08-C is accompanied by a large dust extinction, $A_{\rm V,neb} = 1.5\pm0.6$. This is consistent with a very young, possibly embedded stellar population.}

Taking the obtained \Xio\ at nominal value, the possibilities in accounting for \Xio\ $\geq26.0$ are a) extremely young or exotic stellar populations with a higher intrinsic \Xio\ than predicted in classic stellar population models; b) non-stellar ionizing source; c) extreme nebular conditions, such as unusually high temperatures $\gtrsim 20,000$~K, with consequent Case~B departures \citep{Katz2024}; \textbf{d) a top-heavy IMF, which can produce \logxio\ values up to 26.1 in metal-free populations \citep{Raiter2010}}. Follow-up work is required to identify evidence of possible non-stellar sources, such as AGN, and to constrain the nebular conditions and the nature of the underlying stellar population in REBELS-08-C. The AGN possibility is particularly intriguing, given that the line ratios [N~II]$\lambda$6584/H$\alpha$ $= 0.33 \pm0.04$ and [O~III]$\lambda$5007/H$\beta = 3.84\pm0.55$ of REBELS-08-C place it just barely in the AGN zone of the BPT diagram \citep{Baldwin1981}, based on the \cite{Kewley2001} boundaries (although, see e.g., \citet{Shapley2005, Liu2008, Steidel2014, Strom2017}, suggesting that these demarcations may not be applicable for identifying AGN at high $z$.)

Previous works have reported similar values of \Xio\ exceeding the theoretical threshold $>26.0$, e.g. individual measurements of $>26.4$ for strong Ly$\alpha$ emitters at $z\sim6$ \citep{Ning2023, Simmonds2023}, or even a sample average of 26.3 at $z = 4-5$ for 35 UV-faint Ly$\alpha$ emitters \citep{Maseda2020}. The latter could be explained with a very young, $\leq 3$~Myr, metal-poor, $Z < 0.4~\rm Z_\odot$, stellar population. Our findings show that within more massive, metal-rich, and UV-luminous galaxies such as those in the REBELS sample, there may exist extreme star-forming regions similar in properties to the dwarf starbursts thought to dominate reionization, though with likely higher stellar masses and metallicities.

\subsection{Escaping Ionizing Radiation and Ionized Bubbles of REBELS-IFU Galaxies}
\label{sec:bubbles}
With our global measurements of the intrinsic \Ha\ and \Hb\ luminosities providing an estimate of $Q_{\rm H^0}$, we compute the Strömgren radii of these galaxies, i.e. upper limits on the radii of ionized bubbles they can drive. We note that these galaxies are unlikely to fill their Strömgren radii, in particular those with high \Xio\ and thus dominated by young stellar populations. We assume the cosmic hydrogen number density given by $n_{\rm H} = 1.9 \times10^{-7}~(1+z)^3~\rm cm^{-3}$  \citep{Planck2018} and recombination coefficient $\alpha_B = 2.6 \times 10^{-13} \rm~cm^3~s^{-1}$ for $T=10^4 ~\rm K$. The resulting Strömgren radii $R_{\rm S}$ are within the range $2.7 - 4.9 \rm$ proper Mpc (pMpc). 

The Strömgren radii are upper limits on the sizes of H~II bubbles these galaxies can drive because in deriving them, we assume that all generated LyC photons within the galaxies go on to ionize the IGM, i.e. that $f_{\rm esc,LyC} = 100\%$. This is of course not the case, and in fact the opposite of our assumption of $f_{\rm esc,LyC} = 0\%$ in deriving \Xio\,, because these galaxies show bright emission lines and thus internal reprocessing of ionizing photons, in contrast to, e.g., remnant leakers \citep{Baker2025}. However, if $f_{\rm esc, LyC}$ were instead $\sim2\%$, as we estimate below, these radii would decrease to $\sim1-2$~pMpc. The LyC escape fractions of our sources are therefore key to constrain, in order to understand the potential impact of REBELS-IFU and similar galaxies in ionizing the IGM.

In absence of direct LyC measurements, Ly$\alpha$ is one of the strongest indirect tracers of LyC escape \citep{Verhamme2015, Izotov2021, Flury2022b, SaldanaLopez2025}. \cite{Endsley2022lya} present Ly$\alpha$ observations for three REBELS-IFU galaxies, REBELS-14, 15, and 39. With our spectroscopic \Ha\ observations, corrected for dust, we now have a new constraint on $Q_{\rm H^0}$ and therefore on the Ly$\alpha$ escape fractions $f_{\rm esc,Ly\alpha}$, obtaining $5.2_{-0.7}^{+1.0}\%, 0.9_{-0.2}^{+0.3}\%$, and $3.2_{-0.7}^{+1.1}\%$, respectively. We caution that the estimate for REBELS-14 is particularly uncertain due to the lack of \Ha\ coverage, where we instead use \Hb. These escape fractions are reasonably consistent with the SED-based estimates of \cite{Endsley2022lya}: $5.0_{-1.4}^{+2.9}\%, 0.4_{-0.1}^{+0.2}\%$ and $1.7_{-0.3}^{+0.4}\%$, respectively. The LyC escape fractions of these galaxies are $<5\%$, since Ly$\alpha$ escape fractions are known to be higher than LyC ones \citep{Dijkstra2016, Begley2024, Choustikov2024}. This is because Ly$\alpha$ photons can more readily escape due to resonant scattering. For example, in a large sample of star-forming galaxies at $z \sim 4-5$ \cite{Begley2024} find $f_{\rm esc, LyC} = 0.15 \times f_{\rm esc, Ly\alpha}$. 

We can gain additional insights into the LyC escape of these sources from the results of \cite{Hayes2023}, who model the Ly$\alpha$ emission of 23 galaxies at $z > 6$, including the three REBELS-IFU galaxies above. Combining low-$z$ Ly$\alpha$ spectral templates with IGM damping-wing absorption, they fit the observed Ly$\alpha$ EWs and velocity offsets, to infer the galaxies' distances to an H~I screen, i.e. ionized bubble radii. For REBELS-14, 15, and 39, they find ionized bubble radii of $R_{\rm B} = 1.7_{-0.6}^{+0.4}$~pMpc, $0.9_{-0.4}^{+0.7}$~pMpc, and $1.8_{-0.6}^{+0.4}$~pMpc, respectively. From comparing our new Strömgren radius estimates to these bubble sizes, we can derive the LyC escape fraction as $f_{\rm esc, LyC} = (R_{\rm B}/R_{\rm S})^3$. We obtain $f_{\rm esc, LyC} = 3.1_{-1.6}^{+2.4}\%$, $0.4_{-0.3}^{+0.7}\%$, and $2.7_{-1.4}^{+2.3}\%$, respectively, with a weighted average $0.8_{-0.4}^{+0.6}\%$. These toy estimates are likely upper limits, since we ignore clumping, temporal evolution of the ionization front, as well as anisotropy of LyC escape.

\begin{figure}
\includegraphics[width=\columnwidth]{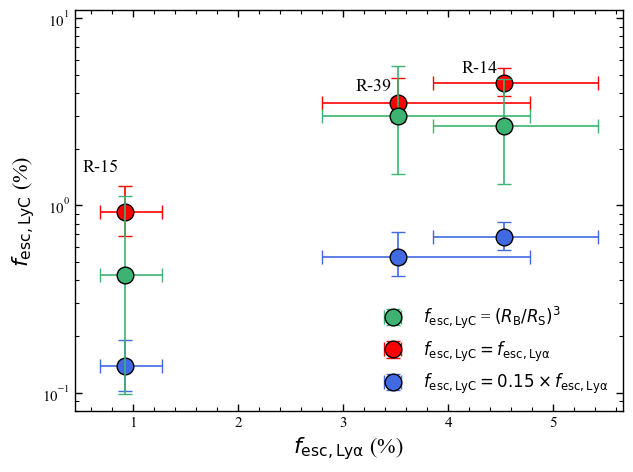} 
\caption{Ly$\alpha$ and LyC escape fractions of three REBELS-IFU galaxies. $f_{\rm esc, Ly\alpha}$ values are computed from the Ly$\alpha$ fluxes reported by \cite{Endsley2022lya} and our de-reddened \Ha\ luminosities. For comparison, we overplot LyC escape fraction predictions based on Ly$\alpha$: upper limits $f_{\rm esc, LyC} \leq f_{\rm esc, Ly\alpha}$ (red) and $f_{\rm esc, LyC} = 0.15 \times f_{\rm esc, Ly\alpha}$ \citep{Begley2024} (blue). Finally, we show the fiducial $f_{\rm esc, LyC}$ values obtained as $(R_{\rm B}/R_{\rm S})^3$ (green), where $R_{\rm B}$ are the ionized bubble radii \cite{Hayes_2023} estimate from the observed Ly$\alpha$ properties of these objects, and $R_{\rm S}$ are Strömgren radii we obtain as described in Section~\ref{sec:bubbles}. }
\label{fig:fesc}
\end{figure}

We compare these fiducial LyC escape fractions to Ly$\alpha$ escape fractions in Figure~\ref{fig:fesc}, where we additionally show predictions for the LyC escape fraction based on Ly$\alpha$, assuming $f_{\rm esc, LyC} = f_{\rm esc, Ly\alpha}$ and $f_{\rm esc, LyC} = 0.15\times f_{\rm esc, Ly\alpha}$ \citep{Begley2024}. We see that the LyC escape fractions of REBELS-14 and 39 are likely higher than $0.15\times f_{\rm esc, Ly\alpha}$, and consistent within error bars with this prediction for REBELS-15. All three objects are consistent with the two escape fractions being equal, within the large uncertainties. This shows that either the LyC escape fractions are somewhat overestimated, or perhaps both Ly$\alpha$ and LyC photons escape along clear sightlines with too little H~I or dust to scatter Ly$\alpha$, as in a picket-fence geometry \citep{Heckman2001, Heckman2011, Jaskot2019, Gazagnes2020}. Indeed, \cite{Rowland2025_DLA} find damped Ly$\alpha$ absorption, tracing significant H~I reservoirs, in all but these three REBELS-IFU galaxies. \textbf{On the other hand, REBELS-14, 15, and 39 show substantial Ly$\alpha$ velocity offsets of $165-324 \rm~km~s^{-1}$ \citep{Endsley2022lya}, indicative of significant H~I column densities, $\log(N_{\rm HI}/\rm cm^{-2}) \sim 19-20$ \citep{Hashimoto2015}, and not consistent with direct LyC/Ly$\alpha$ escape via clear sightlines.}

\textbf{The average LyC escape fraction of REBELS-14, 15, and 39, is $\sim 1\%$, providing a benchmark for the ionizing photon escape in the broader REBELS sample. Although this value is likely an upper limit, considering that Ly$\alpha$ is not detected in half of the probed REBELS galaxies \citep{Endsley2022lya}.} LyC escape fractions of $0.4-3\%$ are consistent with multivariate predictions of an average $f_{\rm esc, LyC} \sim 5\%$ at $z \gtrsim 6$ \citep{Jaskot2024}, and in fact in perfect agreement with the Attenuation-Free Model, which, at our sample mean $z=6.9$, predicts $2-5\%$ \citep{Ferrara2025}. It is interesting to highlight that ALMA observations show the bulk of the REBELS-IFU galaxies to be quite dusty \citep{Inami2022, Algera2025}, demonstrating that even in these dust-rich systems, tangible LyC escape is still possible.

\begin{figure*}
\includegraphics[width=\textwidth]{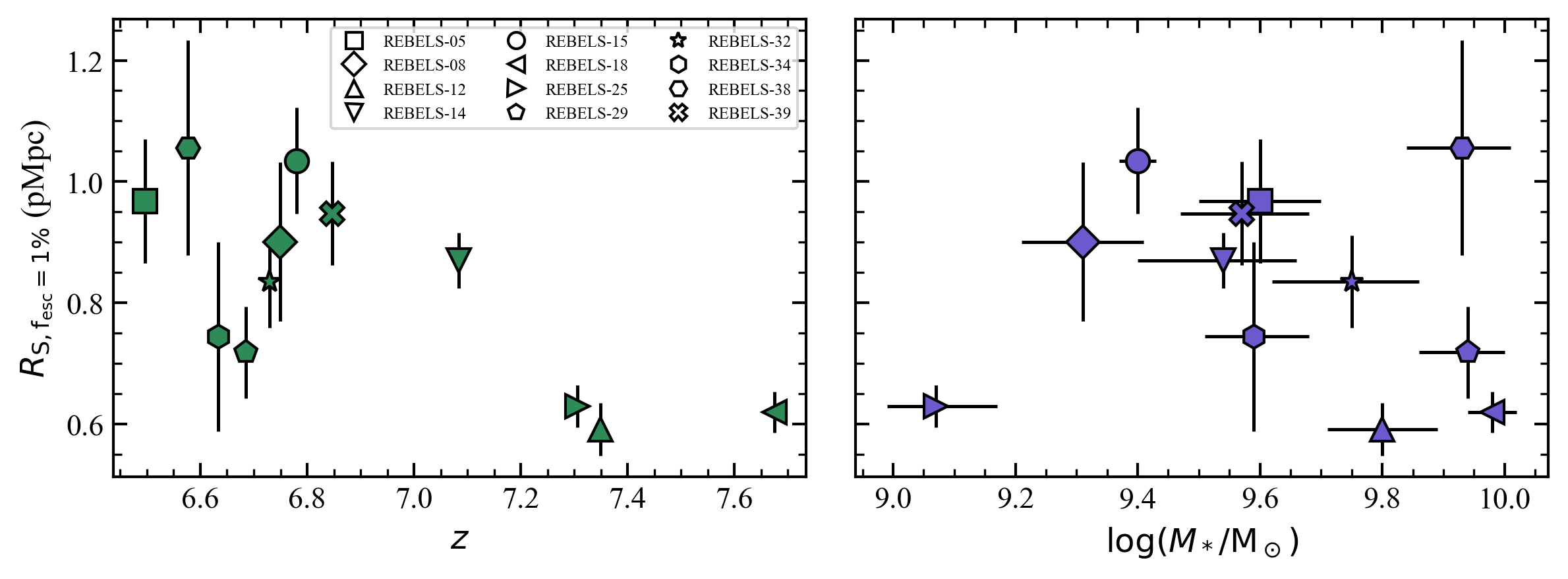} 
\caption{Estimates of the radii of ionized bubbles driven by the galaxies in our sample as a function of redshift, assuming $f_{\rm esc,LyC} =1\%$. The Strömgren radii are computed from dust-corrected \Ha\ or \Hb\ ($z\geq7.0$) luminosities, assuming $n_H = 1.9 \times10^{-7}~(1+z)^3~\rm cm^{-3}$ and recombination coefficient $\alpha_B = 2.6 \times 10^{-13} \rm~cm^3~s^{-1}$ for $T=10^4 ~\rm K.$ Left: $R_{\rm S}$ vs. redshift. Right: $R_{\rm S}$ vs. stellar mass.}
\label{fig:bubbles}
\end{figure*}

If we now compute the ionized bubble radii for the remaining 9 REBELS-IFU galaxies without Ly$\alpha$ measurements from our Strömgren radii and assuming  $f_{\rm esc, LyC} = 1\%$, we obtain $R_{\rm B} = 0.6 -1.1$~pMpc, as shown in Figure~\ref{fig:bubbles} as a function of redshift and stellar mass. So, the REBELS-IFU galaxies show on average global \logxio $\sim 25.4$, $f_{\rm esc, LyC} \sim 1\%$, an emitted ionizing efficiency $\xi_{\rm ion,0}\times f_{\rm esc,LyC} \sim 3 \times 10^{23} \rm~Hz~erg^{-1}$, and drive ionized bubbles of radius $\sim1$~pMpc. 

These resulting ionized bubbles are comparable in size to those previously identified in the literature. For example, the remaining 20 UV-bright galaxies at $z = 6 - 10$, analyzed by \cite{Hayes_2023}, appear to drive bubbles with radii $0.5–2.5$~pMpc, with larger bubbles at lower redshifts. \cite{Endsley2022} find that 10 UV-bright ($M_{\rm UV} \leq -20.4$) Ly$\alpha$-detected galaxies at $z \sim 7$ may generate ionized bubbles with radii $0.7-1.1$~pMpc. \cite{Lu2024} perform large-scale \texttt{21cmFAST} \citep{Mesinger2011} reionization simulations and find that in their rapid reionization model, driven by rare luminous galaxies, the ionized bubble size distribution peaks at $R_{\rm B} > 20$~cMpc or $\sim 2.5$~pMpc, as early as $\bar{x}_{\mathrm{HI}} = 0.7$. \textsc{THESAN} simulation results are consistent with this, resulting in $R_{\rm B} \sim 20$~cMpc for $M_{\rm UV} < -21$~mag galaxies at $z = 7$ \citep{Neyer2024}. 

So, UV-bright galaxies, like those in the REBELS-IFU sample, likely carve out $1-2$~pMpc ionized bubbles, in line with both observations and simulations. Despite their rarity and low LyC escape fractions, they may play a key role in reionization by driving the growth of the largest bubbles, pre-clearing their immediate environments of neutral hydrogen and shaping the topology of reionization.

\section{Conclusions}
\label{sec:conclusions}
We have analyzed the JWST NIRSpec-IFU spectra of 12 UV-bright $z= 6.5 - 7.7$ galaxies from the REBELS program, to measure their integrated and spatially resolved ionizing photon (LyC) production efficiencies \Xio. Our global measurements reveal \Xio\ values largely consistent with the canonical value of \logxio = 25.3 in most of our galaxies, ranging from $25.19\pm0.11$ to $25.61\pm0.11$. With the sample of 25 individual, resolved star-forming clumps within these galaxies, we probe the local LyC production efficiencies. We find that the local \Xio\ values show a much larger scatter of almost 2 dex, ranging between $24.52\pm0.21$ and $26.18\pm0.61$. A possibly extreme value of \logxio $> 26.0$ is identified in one clump within REBELS-08, diluted in the lower global value of $25.58\pm0.17$. This shows that the production of ionizing photons may be inhomogeneous within massive EoR galaxies.

To identify the physical properties driving \Xio\ values in our sample, we relate galaxy-scale and clump-scale \Xio\ to $z$, line and UV luminosities, $\beta$, $\rm EW_0(H\beta)$, $\rm sSFR_{\rm H\beta}$, $\Sigma_{\mathrm{SFR, H\beta}}$, $\log(U)$, $12 + \log(\rm O/H)$, and $M_{*,\rm tot}$. We find a strong correlation of the global \Xio\ with global EW(H$\beta$) and $\rm sSFR_{\rm H\beta}$, as well as an anti-correlation with $M_{*,\rm tot}$, consistent with previous studies. We recover these correlations on the scale of individual clumps as well, with the resolved $\rm EW_0(H\beta)$-\Xio\ relation being significantly steeper than the global one, suggesting a tight physical coupling between local stellar population age and the \Xio\ that is diluted on galaxy scales. In addition, we find the local \Xio\ to correlate with the local $\Sigma_{\mathrm{SFR, H\beta}}$. 
Our analysis thus shows that younger, less massive, and more densely star-forming stellar populations have higher ionizing photon production efficiencies.

We additionally report the upper limits on the radii of ionized bubbles that REBELS-IFU galaxies can drive, estimated as Strömgren radii with the assumption $f_{\rm esc,LyC} = 100\%$, finding $R_{\rm S} \sim 3 - 5$~pMpc. For three of our galaxies, we use the bubble size measurements $R_{\rm B}$ by \cite{Hayes_2023} based on the observed Ly$\alpha$ properties. Comparing these two estimates, we find an average $f_{\rm esc, LyC} \sim 1\%$ for the REBELS-IFU galaxies, and assume this value to estimate $R_{\rm B}$ for the rest of the sample, finding $0.6 -1.1$~pMpc. So, the UV-bright, massive REBELS galaxies may generate ionized bubbles $\sim 1$~pMpc in radius, consistent both with observations of other Ly$\alpha$-detected UV-bright galaxies and reionization simulations. 

Our findings thus shed new light on cosmic reionization. The large variability of \Xio\ within galaxies, as well as possibly extreme values of \Xio\ in individual star-forming regions, should be accounted for in future modeling efforts. Instead of a single \Xio\ value, a distribution of \Xio\ within galaxies needs to be adopted, in addition to the known relationships between \Xio\ and local physical properties.

With even a low number density and modest LyC escape fractions, e.g. $\sim 1\%$, UV-bright, massive galaxies could contribute significantly to ionizing their environments. In particular, because they reside in large halos and overdense regions, their output can drive early bubble growth. While reionization likely remained dominated by numerous fainter galaxies, the massive ones could light the path by quickly ionizing their immediate surroundings. 

\section{Acknowledgements}
\textbf{We thank the referee for their constructive comments, which have helped us to improve the manuscript.} This work was funded by ERC consolidator grant CPI-24-181. MS acknowledges support from the European Research Commission Consolidator Grant 101088789 (SFEER), from the CIDEGENT/2021/059 grant by Generalitat Valenciana, and from project PID2023-149420NB-I00 funded by MICIU/AEI/10.13039/501100011033 and by ERDF/EU. MA is supported by FONDECYT grant number 1252054, and gratefully acknowledges support from ANID Basal Project FB210003 and ANID MILENIO NCN2024\_112.  RB acknowledges support from an STFC Ernest Rutherford Fellowship [grant number ST/T003596/1]. P. Dayal warmly acknowledges support from an NSERC discovery grant (RGPIN-2025-06182). 

This work is based on observations made with the NASA/ESA/CSA James Webb Space Telescope. The data were obtained from the Mikulski Archive for Space Telescopes at the Space Telescope Science Institute, which is operated by the Association of Universities for Research in Astronomy, Inc., under NASA contract NAS 5-03127 for JWST. These observations are associated with programs GO 1626 and GO 2659.

\appendix
\section{Clump Extraction}
The complete set of nebular 2D models and corresponding clump extraction maps are shown in Figure~\ref{fig:2dfits_app}.

\setcounter{figure}{0}
\renewcommand{\thefigure}{A\arabic{figure}}

\begin{figure*}[htbp]
    \centering
    \includegraphics[width=0.75\textwidth]{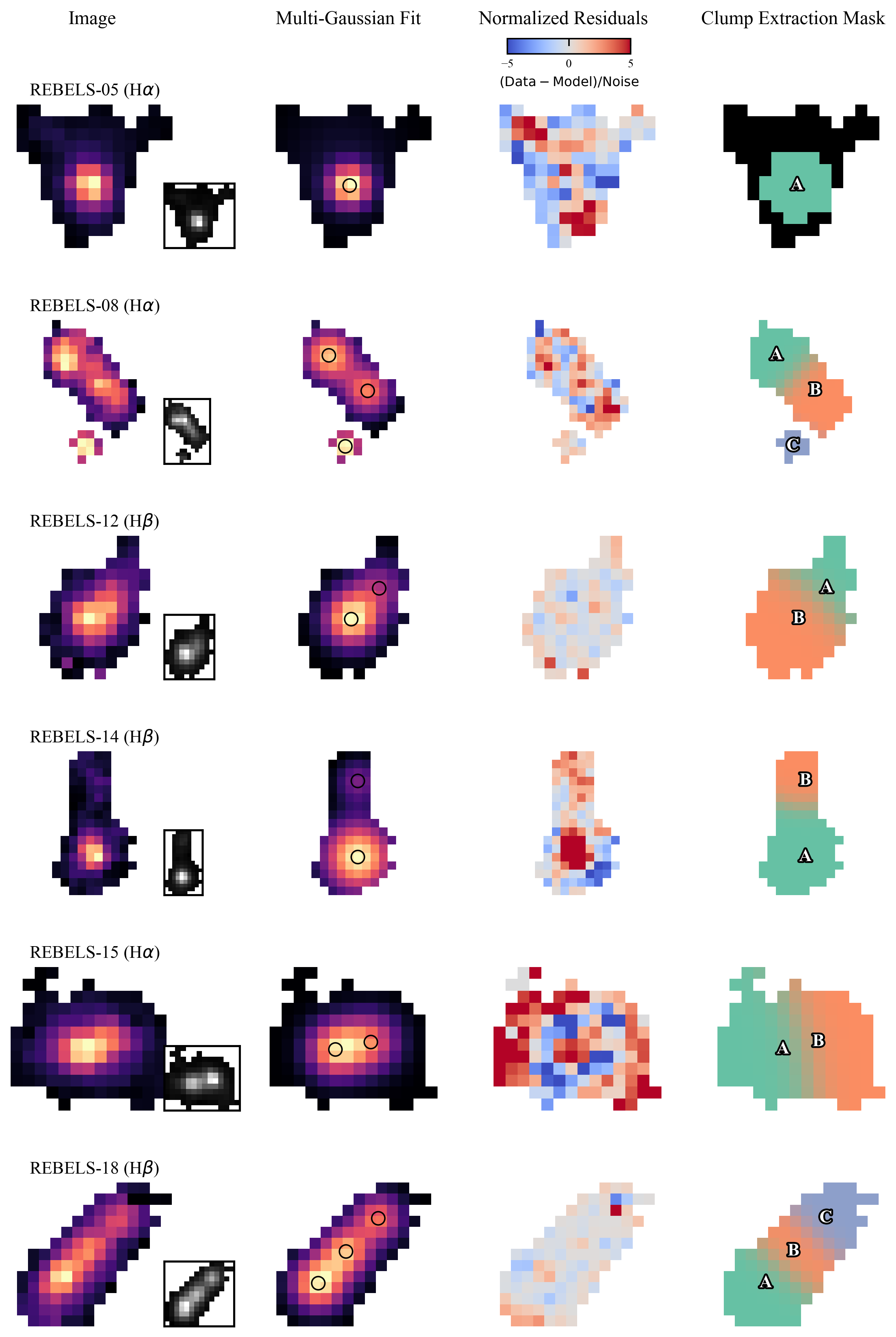}
    \caption{Clump extraction results for the REBELS-IFU sample. From left to right, the panels show: continuum-subtracted \Ha\ or \Hb\ image used for clump identification, with an inset of a UV continuum image ($1250-2600$~\AA); 2D multi-Gaussian model fit to this image; normalized residuals, defined as (data - model)/$\sigma$; the final mask used to extract the spectra of each clump from the IFU cube. The best-fit clump centroids are indicated with black circles in the second panel and labeled A-C in the last one.}
    \label{fig:2dfits_app}
\end{figure*}

\setcounter{figure}{0}
\begin{figure*}[htbp]
    \centering
    \includegraphics[width=0.75\textwidth]{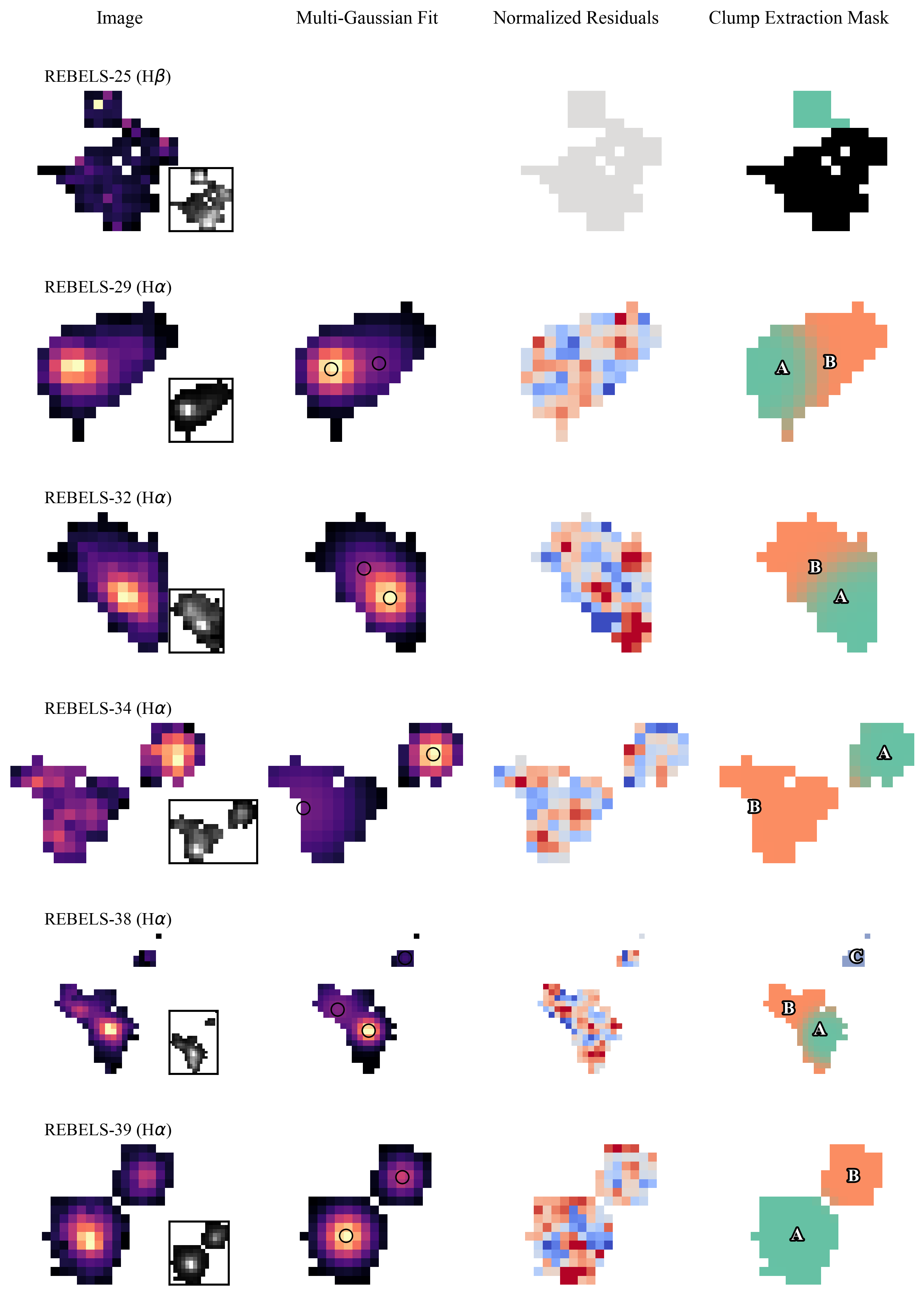}
    \caption{Continued.}
    \label{fig:2dfits_app2}
\end{figure*}

\section{Broader Comparison of Measured \Xio\ to Previous Studies}
\textbf{We present a comparison of our obtained \Xio\ values to a more general compilation of earlier measurements \citep{Stark2015, Schaerer2016, Bouwens2016, Matthee2017, Stark2017, Shivaei2018, Emami2020, Nanayakkara2020, Endsley2021, Atek2022, Stefanon2022, Matthee2023, Castellano2023, Simmonds2023, Hsiao2024, AlvarezMarquez2024, Llerena2024_Xi, Calabro2024, PrietoJimenez2025, AlvarezMarquez2025, Pahl2025}, across stellar masses in Figure~\ref{fig:app_lit}. }

\setcounter{figure}{0}
\renewcommand{\thefigure}{B\arabic{figure}}

\begin{figure*}[htbp]
    \centering
    \includegraphics[width=0.9\textwidth]{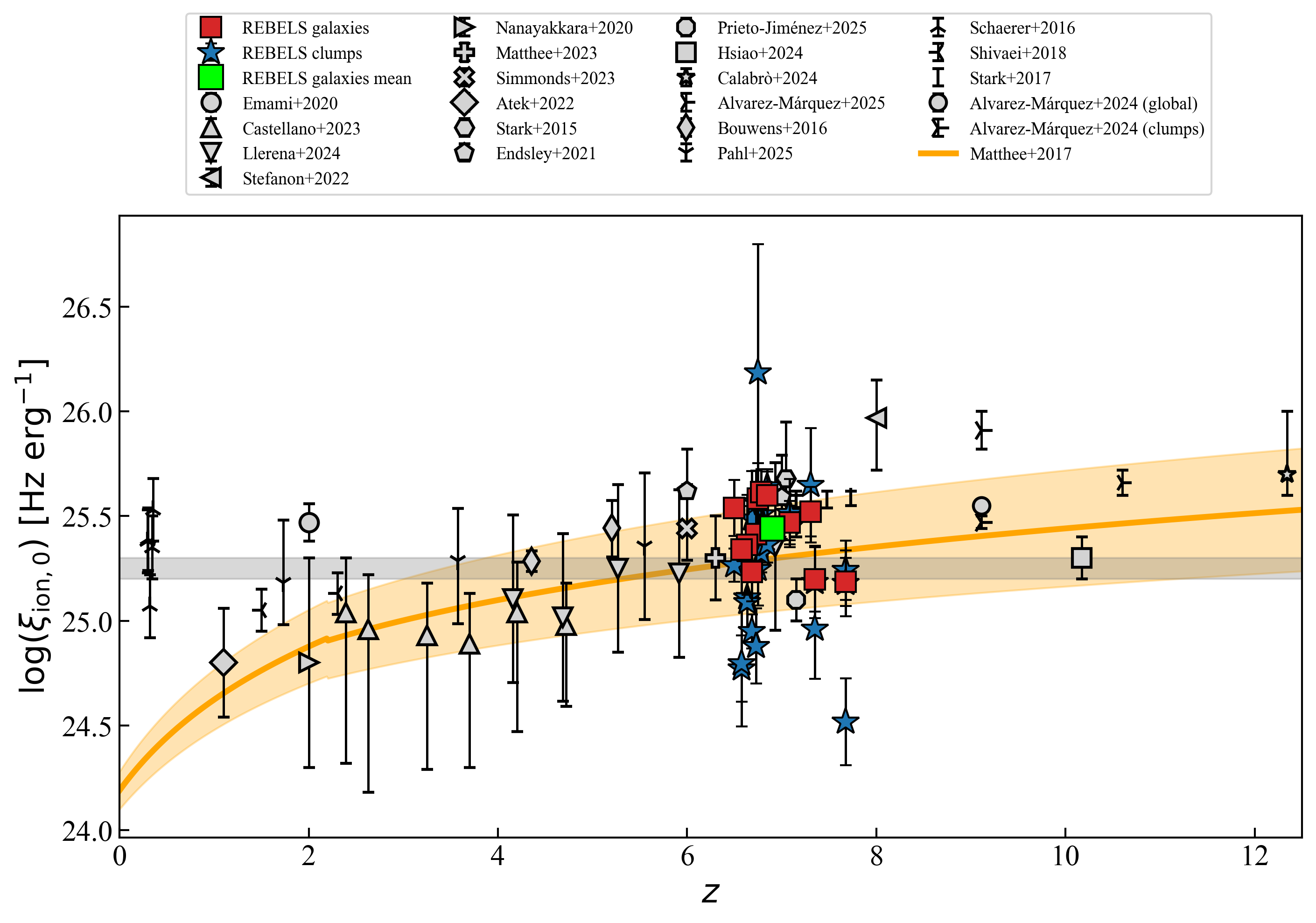}
    \caption{Same as Figure~\ref{fig:litcomp}, but including stellar masses lower than $10^{8.7}~\rm M_{\odot}.$}
    \label{fig:app_lit}
\end{figure*}

\end{document}